\documentclass[aps, prb, showpacs, superscriptaddress, twocolumn]{revtex4} 
\usepackage{graphicx}
\usepackage{dcolumn}
\usepackage{amsmath}
\usepackage{amssymb}
\usepackage{color}
\usepackage{bm}

\newcommand{\bvec}[1]{\mbox{\boldmath $#1$}}

\begin{document}
\preprint{HEP/123-qed}
\title{Dimensional reduction and its breakdown in the driven random field $O(N)$ model}
\author{Taiki Haga}
\affiliation{Department of Physics, Kyoto University, Kyoto 606-8502, Japan}
\email[]{haga@scphys.kyoto-u.ac.jp}
\date{\today}

\begin{abstract}
The critical behavior of a random field $O(N)$ model driven at a uniform velocity is investigated near three dimensions at zero temperature.
From intuitive arguments, we predict that the large-scale behavior of the $D$-dimensional driven random field $O(N)$ model is identical to that of the $(D-1)$-dimensional pure $O(N)$ model.
This is an analogue of the dimensional reduction property of equilibrium cases, which states that the critical exponents of $D$-dimensional random field models are identical to those of $(D-2)$-dimensional pure models.
However, the dimensional reduction property breaks down in low enough dimensions due to  the presence of multiple meta-stable states.
By employing the non-perturbative renormalization group approach, we calculate the critical exponents of the driven random field $O(N)$ model in the first order of $\epsilon=D-3$ and determine the range of $N$ in which the dimensional reduction breaks down.
\end{abstract}

\pacs{11.10.Hi, 05.60.-k, 75.10.Nr}

\maketitle

\section{Introduction}

The effect of quenched disorder on the large-scale structure of interacting systems is still unknown.
The difficulties of disordered systems come from the fact that the competition between disorder and interaction leads to multiple meta-stable states, which are local minima of the Hamiltonian.
The presence of such meta-stable states can significantly affect the critical behaviors and dynamical properties of the system.
One of the most remarkable phenomena associated with meta-stable states is the failure of the so-called ``dimensional reduction'' property in the random field spin models. 
Standard perturbation theory predicts that the critical behaviors of $D$-dimensional random field spin models are the same as those of $(D-2)$-dimensional pure spin models. \cite{Aharony,Parisi}
However, it is known that this dimensional reduction breaks down in low enough dimensions due to a non-perturbative effect associated with multiple meta-stable states.
For example, it predicts that the lower critical dimension of the random field Ising model (RFIM) is three because that of the pure Ising model is one.
This result obviously contradicts the fact that the three-dimensional (3D) RFIM exhibits long-range order (LRO) at weak disorder. \cite{Bricmont}

A promising theoretical approach for describing the breakdown of the dimensional reduction is functional renormalization group (RG) theory. \cite{Fisher-85-1,Fisher-86,Balents-93}
In this formalism, one follows the RG flow of a whole function of a renormalized cumulant for the disorder.
For a particular range of parameters, the renormalized cumulant corresponding to a fixed point exhibits a cusp as a function of the field.
Such non-analytic behavior is a consequence of the presence of multiple meta-stable states and leads to the breakdown of the dimensional reduction.
The range of the parameters for which the dimensional reduction fails has been determined for the RFIM and the random field $O(N)$ model (RF$O(N)$M). \cite{Feldman-02,Tarjus-08,Tissier-12}
For example, the critical dimension above which the dimensional reduction recovers is estimated as $D_{\mathrm{DR}} \simeq 5 $ for the RFIM.
This is consistent with numerical simulations in which the critical exponents of the four and five-dimensional RFIM are calculated. \cite{Fytas-16-1,Fytas-16-2,Fytas-17}

A more challenging problem concerning the effect of disorder, is to understand the critical behavior of disordered systems driven out of equilibrium in the presence of an external force.
It is still poorly understood how the interplay between the quenched disorder and driving force affects the large-scale structure of the system.
One of the well-studied examples is the driven vortex lattices in dirty superconductors. \cite{Blatter}
In such systems, vortex lines are driven by the Lorentz force in the presence of a random pinning potential.
The transport properties associated with the collective dynamics of the vortices have been investigated experimentally. \cite{Charalambous,Bhattacharya,Yaron,Safar}

The most remarkable phenomena in the driven vortex lattices are {\it depinning } and {\it dynamical reordering transitions}.
For a driving force lower than a certain critical value, the lattice is pinned by disorder and the average velocity is zero.
As the driving force becomes large enough to overcome the resistance of the pinning force, the lattice starts to move.
This transition from a pinned state to a moving state is known as a depinning transition, in which the average velocity corresponds to an order parameter.
In recent decades, there have been notable theoretical developments in the depinning transition of elastic systems driven in a disordered medium. \cite{Fisher-85-2,Narayan,Chauve,LeDoussal-02}
The critical exponents near the transition point and the force-velocity characteristics at finite temperature are obtained from the functional RG analysis.

A vortex lattice driven at large velocities becomes more ordered than that for small driving velocities. 
The intuitive explanation is that at large driving velocities the random forces that each vortex experiences vary rapidly and the inhomogeneity is reduced.
Thus, as the driving velocity increases, a phase transition from a disordered phase to an ordered phase takes place.
This nonequilibrium phase transition is called a dynamical reordering transition. \cite{Koshelev,Ryu,Dominguez}
The depinning and dynamical reordering transitions can be observed in a wide variety of systems, such as charge density waves \cite{Gruner} and colloids \cite{Korda} driven over inhomogeneous substrates.

The vortex lattice system in a dirty superconductor is equivalent to the random field XY model at weak disorder.
In this sense, the vortex lattice can be considered as a disordered system with ``Abelian'' symmetry.
On the other hand, there are some examples of random ``non-Abelian'' systems, which include liquid crystals confined in random porous media. \cite{Feldman-00-LC,Bellini,Petridis}
The universality class of critical phenomena crucially depends on the symmetric nature of the system.
Therefore, it is an intriguing question to ask what kinds of phase transitions occur when random non-Abelian systems are driven out of equilibrium.

To resolve this problem, it is a natural starting point to introduce a nonequilibrium counterpart of the dimensional reduction property.
In this study, we first attempt to establish a dimensional reduction which relates the nonequilibrium steady states of driven disordered systems to the equilibrium states of lower dimensional pure systems.
This implies that the critical exponents of the dynamical reordering transitions in driven disordered systems are the same as those of the equilibrium phase transitions in the corresponding pure systems.

The strategy of this study is as follows:
(i) We introduce a simple model of non-Abelian systems, the driven random field $O(N)$ model (DRF$O(N)$M), which is the RF$O(N)$M driven at a uniform and steady velocity.
This model exhibits a dynamical reordering transition above three dimensions.
(ii) From intuitive arguments, we derive a novel type of dimensional reduction property which states that the critical exponents of the $D$-dimensional DRF$O(N)$M at zero temperature are identical to those of the $(D-1)$-dimensional pure $O(N)$ model.
However, as in equilibrium, it can break down in low enough dimensions due to a non-perturbative effect associated with multiple meta-stable states.
(iii) By employing the non-perturbative renormalization group (NPRG) approach, we investigate the critical behavior of the model at zero temperature.
(iv) From the non-analytic behavior of the renormalized disorder correlator, we determine the region in the parameter space wherein the dimensional reduction breaks down.

According to the above strategy, this paper is organized as follows.
In Sec.~\ref{sec:Model}, we define the DRF$O(N)$M and discuss the lower critical dimension of this model.
We also review the phase structure of the RF$O(N)$M.
In Sec.~\ref{sec:DR}, we introduce a dimensional reduction property for the DRF$O(N)$M from naive arguments, as well as its limitations.
In Sec.~\ref{sec:NPRG}, the formalism of the NPRG method is developed for our model.
We derive the RG equation for the renormalized cumulant of the random field, which is a set of coupled nonlinear partial differential equations.
In Sec.~\ref{sec:Results}, the critical exponents are calculated as functions of $N$ in the first order of $\epsilon=D-3$.
We show that the dimensional reduction breaks down when $2<N<10$ near three-dimensions.
In Sec.~\ref{sec:Summary}, the results are summarized.

\section{Model}
\label{sec:Model}

Let $\bvec{\phi} (\bvec{r})=(\phi^1(\bvec{r}),...,\phi^N(\bvec{r}))$ be an $N$-component real vector field.
The Hamiltonian of the $O(N)$ model with a quenched random field $\bvec{h}(\bvec{r})$ is given by
\begin{equation}
H[\bvec{\phi};\bvec{h}]=\int d^D \bvec{r} \biggl[ \frac{1}{2} K |\nabla \bvec{\phi}|^2+U(\rho) - \bvec{h} \cdot \bvec{\phi} \biggr],
\label{Hamiltonian}
\end{equation}
where $\rho=|\bvec{\phi}|^2/2$ is the field amplitude and $U(\rho)=(\lambda_0/2) (\rho-\rho_{0})^2$ is a local interaction potential.
The random field obeys a mean-zero Gaussian distribution with
\begin{equation}
\overline{ h^{\alpha}(\bvec{r}) h^{\beta}(\bvec{r'}) } =h_0^2 \delta^{\alpha \beta} \delta(\bvec{r}-\bvec{r'}),
\end{equation}
where the over-bar represents the average over the quenched disorder.
The dynamics are described by
\begin{equation}
\Gamma \bigl( \partial_t \phi^{\alpha}+v \partial_x \phi^{\alpha} \bigr) =- \frac{\delta H[\bvec{\phi};\bvec{h}]}{\delta \phi^{\alpha}}+\xi^{\alpha},
\label{EM}
\end{equation}
where $v$ denotes the uniform time-independent driving velocity, and $\xi^{\alpha}(\bvec{r},t)$ represents the thermal noise that satisfies 
\begin{equation}
\langle \xi^{\alpha}(\bvec{r},t) \xi^{\beta}(\bvec{r'},t') \rangle =2 \Gamma T \delta^{\alpha \beta} \delta(\bvec{r}-\bvec{r'}) \delta(t-t').
\end{equation}
We call this model the driven random field $O(N)$ model (DRF$O(N)$M).
In this study, we consider the case $N \geq 2$.
Since we are interested in the nonequilibrium steady states of this model, in the following, $\langle...\rangle$ denotes the average over the distribution function of the steady state,
\begin{equation}
\langle A[\bvec{\phi}] \rangle \equiv \int \mathcal{D} \bvec{\phi} A[\bvec{\phi}] P_{\mathrm{st}}[\bvec{\phi};\bvec{h}],
\end{equation}
where $P_{\mathrm{st}}$ is the probability distribution function of the steady state for a given realization of the random field. 
The disorder average is given by
\begin{equation}
\overline{ \langle A[\bvec{\phi}] \rangle } \equiv \int \mathcal{D} \bvec{h} \langle A[\bvec{\phi}] \rangle P_{\mathrm{R}}[\bvec{h}],
\end{equation}
where $P_{\mathrm{R}}$ is the distribution function of the random field.

The DRF$O(N)$M describes the relaxation dynamics of ordering systems flowing in a random environment.
An example is liquid crystals flowing in a porous medium.
Recently, the dynamics of liquid crystals confined in a complex geometry has attracted considerable attention, due to not only the fundamental research interest, but also the industrial applications. \cite{Araki,Sengupta}
For liquid crystals in a porous medium, the irregular surface structure of the solid substrate results in symmetry breaking random anchoring, which is similar to the random field in the $O(N)$ model.

We briefly review the equilibrium phase structure of the RF$O(N)$M.
From phenomenological and rigorous arguments, it is shown that the lower critical dimension of the RF$O(N)$M with $N \geq 2$ is four; \cite{Imry,Aizenman} therefore, there is no LRO in three-dimensions.
However, the existence of quasi-long-range order (QLRO), in which the correlation function decays with power-law, is a more subtle problem.
In early theoretical and numerical studies, \cite{Giamarchi-94,Feldman-00,Gingras} it was suggested that the 3D random field XY model (RFXYM), which corresponds to $N=2$, exhibits QLRO at weak disorder.
This phase is called the ``Bragg glass phase'' in the context of the vortex lattice in superconductors. \cite{Menon}
However, recently, more sophisticated renormalization group approaches have been developed, and these studies negated the existence of QLRO in the 3D-RFXYM. \cite{LeDoussal-06,Tissier-06-NPRG,Tissier-06-2-loop}
Therefore, it is believed that the RF$O(N)$M with $N \geq 2$ neither exhibits QLRO nor LRO in three-dimensions.

We now consider the lower critical dimension of the DRF$O(N)$M, and denote the transverse fluctuation of the order parameter field from a completely ordered state as $\bvec{\phi}^T(\bvec{r})$. 
Its equation of motion at zero temperature is given by
\begin{equation}
\Gamma \bigl( \partial_t \bvec{\phi}^T+v \partial_x \bvec{\phi}^T \bigr) =  K \nabla^2 \bvec{\phi}^T + \bvec{h}^T,
\end{equation}
where $\bvec{h}^T$ is the transverse component of the random field.
Therefore, if the renormalization of the random field is ignored, the disconnected Green's function is given by
\begin{equation}
G^{(T)}_{\mathrm{d}}(\mathbf{q}) = \overline{ \langle \bvec{\phi}^T(\mathbf{q}) \rangle \langle \bvec{\phi}^T(-\mathbf{q}) \rangle }= \frac{ h_0^2}{ K^2 q^4 + \Gamma^2 v^2 q_x^2},
\end{equation}
whose $q$-integral exhibits an infrared-divergence below three dimensions and the lower critical dimension is three.
This implies that, at $D=3+\epsilon$, the DRF$O(N)$M is always in a disordered phase for $v=0$ while it exhibits a dynamical reordering transition at sufficiently large $v$.
In particular, at $D=3$ and $N=2$, one may expect the existence of a QLRO phase and the Kosterlitz-Thouless (KT) transition \cite{KT,Nelson} from the analogy of the two-dimensional (2D) pure XY model.
In fact, it is suggested from the one-loop perturbative renormalization group analysis that elastic lattices driven in a random potential exhibit an anisotropic QLRO at weak disorder. \cite{Giamarchi-96,LeDoussal-98,Balents-98}
Furthermore, the KT-like transition is also observed in numerical simulations of the 3D-DRF$O(2)$M. \cite{Haga}
However, since the analysis for this case requires a more careful treatment than that given in this study, the detailed investigation will be presented elsewhere.

\section{Dimensional reduction}
\label{sec:DR}

In this section, we introduce a dimensional reduction property for the DRF$O(N)$M.
At zero temperature, Eq.~(\ref{EM}) is written as
\begin{eqnarray}
\Gamma(\partial_t \bvec{\phi}+v \partial_x \bvec{\phi}) = K \nabla^2 \bvec{\phi} - U'(\rho) \bvec{\phi} + \bvec{h}(\bvec{r}).
\label{EM_zero_temp}
\end{eqnarray}
A steady state $\bvec{\phi}_{\mathrm{st}}(\bvec{r})$ satisfies the following equation:
\begin{eqnarray}
\Gamma v \partial_x \bvec{\phi} = K \nabla^2 \bvec{\phi} - U'(\rho) \bvec{\phi} + \bvec{h}(\bvec{r}).
\label{EM_st}
\end{eqnarray}
In the large length scale, the longitudinal elastic term $K \partial_x^2 \phi$ is negligible compared to the advection term $v \partial_x \phi$.
Thus, Eq.~(\ref{EM_st}) becomes
\begin{eqnarray}
\Gamma v \partial_x \bvec{\phi} = K \nabla_{\perp}^2 \bvec{\phi} - U'(\rho) \bvec{\phi} + \bvec{h}(x,\bvec{r}_{\perp}),
\label{EM_st_perp}
\end{eqnarray}
where $\nabla_{\perp}$ is the derivative operator for the transverse directions and $\bvec{r}_{\perp}$ represents the transverse coordinate.
In Eq.~(\ref{EM_st_perp}), if the coordinate $x$ is considered to be a fictitious time and $\bvec{h}(x,\bvec{r}_{\perp})$ as thermal noise, $\bvec{\phi}_{\mathrm{st}}(x,\bvec{r}_{\perp})$ is identical to the dynamical solution for the $(D-1)$-dimensional pure $O(N)$ model with temperature $T_{\mathrm{eff}} = h_0^2/(2\Gamma v)$.
If we assume that all steady states satisfying Eq.~(\ref{EM_st_perp}) contribute with {\it equal weight}, we can conclude that the large-scale behavior of the $D$-dimensional DRF$O(N)$M at zero temperature is identical to that of the $(D-1)$-dimensional pure $O(N)$ model at finite temperature.
This is the dimensional reduction property for the DRF$O(N)$M.

However, as in equilibrium, the dimensional reduction property can break down.
The reason is that each steady state contributes with {\it non-trivial weight} to the averaged quantities.
To clarify this subtle point, we consider the equilibrium case, in particular the RFIM.
The Hamiltonian of the RFIM has many local minima or meta-stable states.
If we assume that all meta-stable states contribute with equal weight, we have the conventional dimensional reduction property \cite{Parisi}; however, such an assumption is incorrect.
In fact, each meta-stable state contributes with the Boltzmann weight $e^{-E/T}$, where $E$ is the energy corresponding to the meta-stable state.
In particular, at sufficiently low temperatures, the dominant contribution is from the ground state.
The same problem arises in nonequilibrium situations.
Let $P_{\mathrm{st}}[\bvec{\phi};\bvec{h},T]$ be the probability distribution function of the steady state for a fixed random field $\bvec{h}$ and temperature $T$.
In the limit $T \to 0$, $P_{\mathrm{st}}[\bvec{\phi};\bvec{h},T]$ has a sharp peak at the ``most probable'' steady state $\bvec{\phi}_{\mathrm{st}}^{*}(\bvec{r})$, which is one of the solutions of Eq.~(\ref{EM_st_perp}).
If the large-scale behavior of $\bvec{\phi}_{\mathrm{st}}^{*}(\bvec{r})$ is different from that of the $(D-1)$-dimensional pure $O(N)$ model, the dimensional reduction breaks down.
It is worth noting that $P_{\mathrm{st}}[\bvec{\phi};\bvec{h},T]$ cannot be expressed in the form $\exp (-E[\bvec{\phi};\bvec{h}]/T)$ because the advection term $v \partial_x \phi$ in Eq.~(\ref{EM}) cannot be cast into the functional derivative of an appropriate potential.

\section{NPRG Formalism}
\label{sec:NPRG}

We employ the NPRG approach \cite{Wetterich} to investigate the large-scale behavior of the DRF$O(N)$M.
This approach enables us to treat interacting systems with large degrees of freedom in a systematic way, such as frustrated magnets, \cite{Delamotte} strongly correlated quantum gases, \cite{Dupuis} reaction-diffusion systems, \cite{Canet-05} and the Kardar-Parisi-Zhang equation. \cite{Canet-10}
The NPRG formalism for disordered systems has been developed in Refs.~\onlinecite{Tarjus-08} and \onlinecite{Tissier-12}.
We extend this formalism to driven disordered systems and derive the RG equations of  the DRF$O(N)$M in the first order of $\epsilon=D-3$.

\subsection{Scale-dependent effective action}

Our starting point is the field-theoretical representation of Eq.~(\ref{EM}).
We introduce the replicated fields $\Phi_a = {}^t(\bvec{\phi}_a, \bvec{\hat{\phi}}_a), \: a=1,..,n$.
Taking the average over the random field leads to the action
\begin{eqnarray}
S[\{ \Phi_a \}] = \sum_a S_1[ \Phi_a ] - \frac{1}{2} \sum_{a,b} S_2[ \Phi_a, \Phi_b ], 
\label{Bare_action}
\end{eqnarray}
where the one and two-replica parts are given by
\begin{eqnarray}
S_1[ \Phi ] = \int_{rt} \Bigl[ \Gamma \bvec{\hat{\phi}} \cdot \bigl(  \partial_t \bvec{\phi} + v \partial_x \bvec{\phi} - T \bvec{\hat{\phi}} \bigr) \nonumber \\
+ \bvec{\hat{\phi}} \cdot \bigl\{ -K \nabla^2 \bvec{\phi} + U'(\rho) \bvec{\phi} \bigr\}  \Bigr],
\label{S_1}
\end{eqnarray}
and
\begin{eqnarray}
S_2[ \Phi_a, \Phi_b ] = \int_{rtt'} h_0^2 \: \bvec{\hat{\phi}}_{a,rt} \cdot \bvec{\hat{\phi}}_{b,rt'}.
\label{S_2}
\end{eqnarray}
The derivations of Eqs.~(\ref{S_1}) and (\ref{S_2}) are presented in Appendix \ref{sec:Appendix_action}.
In the following, a superscript with a Greek alphabet letter represents the index of the field component $\alpha,\beta = 1,...,N$ and a subscript with a Roman alphabet letter represents the replica index $a,b = 1,...,n$. 
By introducing source fields $J_a = {}^t(\bvec{j}_a, \bvec{\hat{j}}_a)$, the generating functional is defined as
\begin{eqnarray}
Z[\{ J_a \}]=\int \prod_a \mathcal{D} \Phi_a \exp \biggl[ -S[\{ \Phi_a \}] \nonumber \\
+ \sum_a \int_{rt}  {}^tJ_a \cdot \Phi_a \biggr].
\end{eqnarray}
The effective action is given by a Legendre transformation of $\ln Z[\{ J_a \}]$,
\begin{eqnarray}
\Gamma[\{ \Psi_a \}] = -\ln Z[\{ J_a \}] + \sum_a \int_{rt} {}^tJ_a \cdot \Psi_a,
\label{Full_effective_action}
\end{eqnarray}
where $\Psi_a = {}^t(\bvec{\psi}_a, \bvec{\hat{\psi}}_a)$ and $J_a$ are related by
\begin{eqnarray}
\Psi_a = \langle \Phi_a \rangle = \frac{\delta}{\delta J_a} \ln Z[\{J_a \}]. \nonumber
\end{eqnarray}

The NPRG formalism is based on an exact RG equation for the scale-dependent effective action $\Gamma_k[\{ \Psi_a \}]$, which includes only high-energy modes with momenta larger than the running scale $k$.
As $k$ goes from the cutoff $\Lambda$ to zero, $\Gamma_k$ interpolates between the bare action Eq.~(\ref{Bare_action}) and the full effective action Eq.~(\ref{Full_effective_action}).
To suppress the contribution from the low-energy modes, a mass-like quadratic term is added to the bare action,
\begin{equation}
\Delta S_k[\{ \Phi_a \}] = \frac{1}{2} \sum_{a} \int_{q} {}^t\Phi_a(q) \: \bvec{\mathrm{R}}_k(\mathbf{q}) \: \Phi_a(-q),
\label{Delta_S_k}
\end{equation}
where we have used the notation $q=(\mathbf{q},\omega)$ and $\int_q=\int d^D \mathbf{q} d \omega/(2\pi)^{D+1}$.
A frequency-independent $2N \times 2N$ matrix $\bvec{\mathrm{R}}_k(\mathbf{q})$ is given by
\begin{eqnarray}
\bvec{\mathrm{R}}_k(\mathbf{q}) = R_k(\mathbf{q}) \left(
\begin{array}{ccc}
0 & 1 \\
1 & 0
\end{array}
\right) \otimes \mathbf{I}_N,
\end{eqnarray}
where $\mathbf{I}_N$ is the $N \times N$ unit matrix, which acts on the field component index.
$R_k(\mathbf{q})$ is a cutoff function, which has a constant value proportional to $k^2$ for $q \ll k$ and rapidly decreases for $q > k$.
The explicit form of $R_k(\mathbf{q})$ will be given later.
We also introduce a $2nN \times 2nN$ matrix
\begin{eqnarray}
\hat{\bvec{\mathrm{R}}}_k(\mathbf{q}) = \bvec{\mathrm{R}}_k(\mathbf{q}) \otimes \mathbf{I}_n,
\end{eqnarray}
where $\mathbf{I}_n$ is the $n \times n$ unit matrix, which acts on the replica index.
From the generating functional with the running scale $k$,
\begin{eqnarray}
Z_k[\{ J_a \}]=\int \prod_a \mathcal{D} \Phi_a \exp \biggl[ -S[\{ \Phi_a \}] \nonumber \\
- \Delta S_k[\{ \Phi_a \}] + \sum_a \int_{rt}  {}^tJ_a \cdot \Phi_a \biggr],
\end{eqnarray}
one defines the scale-dependent effective action through a Legendre transformation,
\begin{eqnarray}
\Gamma_k[\{ \Psi_a \}] = -\ln Z_k[\{ J_a \}] + \sum_a \int_{rt} {}^tJ_a \cdot \Psi_a \nonumber \\
- \Delta S_k[\{ \Psi_a \}],
\end{eqnarray}
where $\Psi_a$ and $J_a$ are related by
\begin{eqnarray}
\Psi_a = \langle \Phi_a \rangle = \frac{\delta}{\delta J_a} \ln Z_k[\{J_a \}]. \nonumber
\end{eqnarray}
It can be shown that $\lim_{k \to \infty} \Gamma_k[\{ \Psi_a \}] = S[\{ \Psi_a \}]$ and $\lim_{k \to 0} \Gamma_k[\{ \Psi_a \}] = \Gamma[\{ \Psi_a \}]$.
(See Ref.~\onlinecite{Canet-11} for the general framework of the NPRG formalism for nonequilibrium systems)

\subsection{Exact flow equation for the effective action}

The exact evolution of $\Gamma_k$ is described by Wetterich's equation, \cite{Wetterich}
\begin{equation}
\partial_k \Gamma_k = \frac{1}{2} \mathrm{Tr} \: \partial_k  \hat{\bvec{\mathrm{R}}}_k(\mathbf{q}) \Bigl[ \Gamma_k^{(2)} + \hat{\bvec{\mathrm{R}}}_k(\mathbf{q}) \Bigr]^{-1} ,
\label{RG_Gamma}
\end{equation}
where $\Gamma_k^{(2)}$ is the second functional derivative,
\begin{equation}
(\Gamma_k^{(2)})^{\mu \nu}_{ab}(q,q')=\frac{\delta^2 \Gamma_k}{\delta \Psi^{\mu}_a(q) \delta \Psi^{\nu}_b(q')},
\end{equation}
and $\mathrm{Tr}$ represents an integration over momentum and frequency as well as a sum over the indices of the replica, the field component, and the two conjugate fields $\{ \psi, \hat{\psi} \}$.
Note that $\hat{\bvec{\mathrm{R}}}_k(\mathbf{q})$ in Eq.~(\ref{RG_Gamma}) implicitly contains the delta function of momentum and frequency.

According to Ref.~\onlinecite{Tarjus-08}, $\Gamma_k$ is expanded in an increasing number of free replica sums,
\begin{eqnarray}
\Gamma_k[\{ \Psi_a \}] = \sum_a \Gamma_{1,k}[ \Psi_a ] - \frac{1}{2} \sum_{a,b} \Gamma_{2,k}[ \Psi_a, \Psi_b ] \nonumber \\
+ \frac{1}{3!} \sum_{a,b,c} \Gamma_{3,k}[ \Psi_a, \Psi_b, \Psi_c ] -...,
\label{Gamma_expansion}
\end{eqnarray}
where the multi-replica parts are related to the cumulants of the renormalized random field.
Inserting Eq.~(\ref{Gamma_expansion}) into Eq.~(\ref{RG_Gamma}) leads to the exact flow equations for $\Gamma_{p,k}$.
To express these in a compact form, we define the one-replica propagator with the infrared cutoff, 
\begin{equation}
\mathrm{P}_k[\Psi] = \bigl[ \Gamma_{1,k}^{(2)}[\Psi] + \bvec{\mathrm{R}}_k(\mathbf{q}) \bigr]^{-1}.
\label{Green_function}
\end{equation}
The exact flow equations for $\Gamma_{1,k}[ \Psi ]$ and $\Gamma_{2,k}[ \Psi_1, \Psi_2 ]$ are
\begin{eqnarray}
\partial_k \Gamma_{1,k}[ \Psi ] = \frac{1}{2} \mathrm{tr} \int_{q} \partial_k \bvec{\mathrm{R}}_k(\mathbf{q}) \Bigl[ \mathrm{P}_k[\Psi] \nonumber \\
+ \mathrm{P}_k[\Psi] \Gamma_{2,k}^{(11)}[\Psi,\Psi] \mathrm{P}_k[\Psi] \Bigr],
\label{RG_Gamma_1}
\end{eqnarray}
\begin{eqnarray}
\partial_k \Gamma_{2,k}[ \Psi_1, \Psi_2 ] = -\frac{1}{2} \mathrm{tr} \int_{q} \partial_k \bvec{\mathrm{R}}_k(\mathbf{q}) \Bigl[ \mathrm{P}_k[\Psi_1] \Bigl\{ \nonumber \\
\Gamma_{2,k}^{(20)}[\Psi_1,\Psi_2] - \Gamma_{3,k}^{(110)}[\Psi_1,\Psi_1,\Psi_2]  \nonumber \\
+ \Gamma_{2,k}^{(11)}[\Psi_1,\Psi_2] \mathrm{P}_k[\Psi_2] \Gamma_{2,k}^{(11)}[\Psi_2,\Psi_1] \nonumber \\
+ \Gamma_{2,k}^{(20)}[\Psi_1,\Psi_2] \mathrm{P}_k[\Psi_1] \Gamma_{2,k}^{(11)}[\Psi_1,\Psi_1] \nonumber \\
+ \Gamma_{2,k}^{(11)}[\Psi_1,\Psi_1] \mathrm{P}_k[\Psi_1] \Gamma_{2,k}^{(20)}[\Psi_1,\Psi_2] \Bigr\} \mathrm{P}_k[\Psi_1] \nonumber \\
+\mathrm{perm}(\Psi_1,\Psi_2) \Bigr],
\label{RG_Gamma_2}
\end{eqnarray}
where we have introduced the following notation,
\begin{eqnarray}
\Gamma_{2,k}^{(11)}[\Psi_1,\Psi_2] = \frac{\delta^2 \Gamma_{2,k}[\Psi_1,\Psi_2]}{\delta \Psi_1 \delta \Psi_2}, \nonumber \\
\Gamma_{2,k}^{(20)}[\Psi_1,\Psi_2] = \frac{\delta^2 \Gamma_{2,k}[\Psi_1,\Psi_2]}{\delta \Psi_1 \delta \Psi_1},
\end{eqnarray}
and $\mathrm{tr}$ in Eqs.~(\ref{RG_Gamma_1}) and (\ref{RG_Gamma_2}) represents the sum over the indices of the field component and the two conjugate fields $\{ \psi, \hat{\psi} \}$.
The derivations of Eqs.~(\ref{RG_Gamma_1}) and (\ref{RG_Gamma_2}) are presented in Appendix \ref{sec:Appendix_exact_flow_equations}.
Note that $\Gamma_{p+1}$ appears on the right-hand side of the flow equation for $\Gamma_p$, thus we have an infinite hierarchy of the coupled flow equations.

\subsection{Derivative expansion}

To solve Eqs.~(\ref{RG_Gamma_1}) and (\ref{RG_Gamma_2}), we must introduce an approximation for the functional form of $\Gamma_{p,k}$.
Since we are interested in the large-scale behaviors of the system, we expand the effective action in an increasing number of derivatives of the field and retain only a limited number of terms.

This systematic truncation scheme is called ``derivative expansion''. \cite{Wetterich}
We use the following functional form for the one-replica part $\Gamma_{1,k}$, which includes the first order of the derivative expansion,
\begin{eqnarray}
\Gamma_{1,k}[ \Psi ] = \int_{rt} \Bigl[ X_k \bvec{\hat{\psi}} \cdot \bigl( \partial_t \bvec{\psi} + v_k \partial_x \bvec{\psi} - T_k \bvec{\hat{\psi}} \bigr) \nonumber \\
+ \bvec{\hat{\psi}} \cdot \bigl\{ -Z_{\parallel,k} \partial_x^2 \bvec{\psi} - Z_{\perp,k} \nabla_{\perp}^2 \bvec{\psi} - F_k(\rho) \bvec{\psi}   \bigr\}  \Bigr],
\label{Gamma_1}
\end{eqnarray}
where we have introduced two field renormalization factors $Z_{\parallel,k}$ and $Z_{\perp,k}$ representing the anisotropy due to the driving.
$F_k(\rho)$ is a renormalized local force, which can be written as the derivative of a potential $F_k(\rho)=-U_k'(\rho)$ in the equilibrium case ($v=0$).

The higher-order terms ignored in Eq.~(\ref{Gamma_1}), such as $\bvec{\hat{\psi}} \cdot \bvec{\psi} \nabla^2 \rho$ and $\bvec{\hat{\psi}} \cdot \nabla^4 \bvec{\psi}$, just yield sub-leading contributions to the critical exponents.
More precisely, at $D=3+\epsilon$, the critical exponents have values of $\mathcal{O}(\epsilon)$, but the contributions from these terms are $\mathcal{O}(\epsilon^2)$. 
This fact can be understood as follows.
The first term $\bvec{\hat{\psi}} \cdot \bvec{\psi} \nabla^2 \rho$ modifies the propagator of the longitudinal (massive) mode.
However, as we will explain in Sec.~\ref{sec:RG_equation}, the contribution from the longitudinal mode is negligible in the first order of $\epsilon=D-3$.
The second term $\bvec{\hat{\psi}} \cdot \nabla^4 \bvec{\psi}$ yields a term proportional to $q^4$ in $\Gamma_{1,k}^{(2)}(q)$.
At a critical point, $\Gamma_{1,k}^{(2)}(q)$ is expected to behave as $q^2(q^2+ck^2)^{-\eta/2}$ near $q=0$, where $\eta$ is the anomalous dimension. \cite{Wetterich}
By expanding this expression around $q=0$, one finds that the coefficients of the higher-order terms $\mathcal{O}(q^4)$ are proportional to $\eta$.
As we will show in Sec.~\ref{sec:RG_equation}, $\eta$ has a value of $\mathcal{O}(\epsilon)$.
Therefore, the contribution from the term $\bvec{\hat{\psi}} \cdot \nabla^4 \bvec{\psi}$ is sub-leading compared to that from the term $\bvec{\hat{\psi}} \cdot \nabla^2 \bvec{\psi}$.

The two-replica part $\Gamma_{2,k}$ is given by
 \begin{eqnarray}
\Gamma_{2,k}[\Psi_1,\Psi_2] = \int_{rtt'} \hat{\psi}^{\mu}_{1,rt} \hat{\psi}^{\nu}_{2,rt'} \Delta^{\mu \nu}_k(\bvec{\psi}_{1,rt},\bvec{\psi}_{2,rt'}),
\label{Gamma_2}
\end{eqnarray}
where $\Delta^{\mu \nu}_k(\bvec{\psi}_{1},\bvec{\psi}_{2})$ is the second cumulant of the renormalized random field.
From the rotational symmetry, $\Delta^{\mu \nu}_k(\bvec{\psi}_{1},\bvec{\psi}_{2})$ can be rewritten as
\begin{eqnarray}
\Delta^{\mu \nu}_k(\bvec{\psi}_{1},\bvec{\psi}_{2}) = \Delta_{00,k}(\rho_1, \rho_2, z) \delta^{\mu \nu} + (4 \rho_1 \rho_2)^{-1/2} \bigl[ \nonumber \\
 \Delta_{12,k}(\rho_1, \rho_2, z) \psi_1^{\mu} \psi_2^{\nu} + \Delta_{21,k}(\rho_1, \rho_2, z) \psi_2^{\mu} \psi_1^{\nu} \nonumber \\
+ \Delta_{11,k}(\rho_1, \rho_2, z) \psi_1^{\mu} \psi_1^{\nu} + \Delta_{22,k}(\rho_1, \rho_2, z) \psi_2^{\mu} \psi_2^{\nu} \: \bigr],
\label{Delta_expansion}
\end{eqnarray}
where $z=\bvec{\psi}_{1} \cdot \bvec{\psi}_{2}/\sqrt{4 \rho_1 \rho_2}$ is the cosine of the angle between $\bvec{\psi}_{1}$ and $\bvec{\psi}_{2}$.
In the equilibrium case, it can be written as the derivative of the cumulant of the random potential,
\begin{equation}
\Delta^{\mu \nu}_k(\bvec{\psi}_{1},\bvec{\psi}_{2}) = \partial_{\psi_1^{\mu}} \partial_{\psi_2^{\nu}} V_k(\bvec{\psi}_{1},\bvec{\psi}_{2}). \nonumber
\end{equation}
This cannot be done for the nonequilibrium case.
The bare value is given by $\Delta^{\mu \nu}_{k=\Lambda}(\bvec{\psi}_{1},\bvec{\psi}_{2}) = h_0^2 \delta^{\mu \nu}$.
We ignore the contributions from the higher order cumulants, $\Gamma_{3,k}=\Gamma_{4,k}=...=0$.

It is worth noting the symmetry of the effective action in the absence of the disorder or driving.
In the equilibrium case ($v=0$), it can be shown that the bare action Eqs.~(\ref{S_1}) and (\ref{S_2}) are invariant under a time-reversal transformation, \cite{Canet-11}
\begin{eqnarray}
  \begin{cases}
    t \to -t, & \\
    \psi^{\mu}_a \to \psi^{\mu}_a, & \\
    \hat{\psi}^{\mu}_a \to \hat{\psi}^{\mu}_a - (1/T) \partial_t \psi^{\mu}_a. &
  \end{cases}
  \label{time_reversal_trans_1}
\end{eqnarray}
Since the effective action Eqs.~(\ref{Gamma_1}) and (\ref{Gamma_2}) should be invariant under the same transformation, we can conclude that the temperature is not renormalized, $T_k=T$, in equilibrium.
In contrast, in the presence of a finite velocity $v \neq 0$, Eq.~(\ref{S_1}) is not invariant under Eq.~(\ref{time_reversal_trans_1}) because the advection term $v \partial_x \phi $ cannot be expressed as a functional derivative of a potential.
Thus, in the non-equilibrium case, the temperature can be renormalized, $T_k \neq T$.

Next, we consider a zero random field case $h_0=0$ with a finite driving velocity, and Eq.~(\ref{S_1}) is invariant under a space-time-reversal transformation,
\begin{eqnarray}
  \begin{cases}
    t \to -t, & \\
    x \to -x, & \\
    \psi^{\mu}_a \to \psi^{\mu}_a, & \\
    \hat{\psi}^{\mu}_a \to \hat{\psi}^{\mu}_a - (1/T) \partial_t \psi^{\mu}_a - (v/T) \partial_x \psi^{\mu}_a. &
  \end{cases}
  \label{time_reversal_trans_2}
\end{eqnarray}
This implies that the driving velocity is not renormalized, $v_k=v$.
This result can be easily understood by noting that, in the absence of the random field, the driven system can be mapped into the corresponding equilibrium system through the Galilei transformation.
In contrast, in the presence of the random field, Eq.~(\ref{S_2}) is not invariant under Eq.~(\ref{time_reversal_trans_2}).
Thus, the driving velocity can be renormalized, $v_k \neq v$.

The renormalized parameters in Eq.~(\ref{Gamma_1}) are given by the following functional derivatives:
\begin{eqnarray}
X_k = \frac{1}{\Omega \mathcal{T}} \partial_{i\omega} \Gamma^{(2)}_{1;\hat{\psi}^2 \psi^2}(\mathbf{p},\omega) \bigr|_{\mathbf{p},\omega=0},
\label{Def_X}
\end{eqnarray}
\begin{eqnarray}
X_k v_k = \frac{1}{\Omega \mathcal{T}} \partial_{(-i p_x)} \Gamma^{(2)}_{1;\hat{\psi}^2 \psi^2}(\mathbf{p},\omega) \bigr|_{\mathbf{p},\omega=0},
\label{Def_v}
\end{eqnarray}
\begin{eqnarray}
2X_k T_k = -\frac{1}{\Omega \mathcal{T}} \Gamma^{(2)}_{1;\hat{\psi}^2 \hat{\psi}^2}(\mathbf{p},\omega) \bigr|_{\mathbf{p},\omega=0},
\label{Def_T}
\end{eqnarray}
\begin{eqnarray}
Z_{\parallel,k} = \frac{1}{\Omega \mathcal{T}} \partial_{p_x^2} \Gamma^{(2)}_{1;\hat{\psi}^2 \psi^2}(\mathbf{p},p_x v_k) \bigr|_{\mathbf{p}}, \nonumber \\
Z_{\perp,k} = \frac{1}{\Omega \mathcal{T}} \partial_{p_{\perp}^2} \Gamma^{(2)}_{1;\hat{\psi}^2 \psi^2}(\mathbf{p},p_x v_k) \bigr|_{\mathbf{p}},
\label{Def_Z}
\end{eqnarray}
\begin{equation}
\sqrt{2\rho} F_k(\rho) = - \frac{1}{\Omega \mathcal{T}} \Gamma^{(1)}_{1;\hat{\psi}^1}(\mathbf{p},\omega) \bigr|_{\mathbf{p},\omega=0},
\label{Def_F}
\end{equation}
where we have used the following notation:
\begin{eqnarray}
\Gamma^{(1)}_{1;\hat{\psi}^{\mu}}(\mathbf{p},\omega) &=& \frac{\delta \Gamma_{1,k}[\Psi]}{\delta \hat{\psi}^{\mu}(\mathbf{p},\omega)}, \nonumber \\
\Gamma^{(2)}_{1;\hat{\psi}^{\mu} \psi^{\nu}}(\mathbf{p},\omega) &=& \frac{\delta^2 \Gamma_{1,k}[\Psi]}{\delta \hat{\psi}^{\mu}(\mathbf{p},\omega) \delta \psi^{\nu}(-\mathbf{p},-\omega)}. \nonumber
\end{eqnarray}
$\Omega$ and $\mathcal{T}$ denote the space-time volume.
The functional derivatives in Eqs.~(\ref{Def_X})--(\ref{Def_F}) are evaluated for a uniform field $ \bvec{\psi}_{rt} \equiv {}^t(\sqrt{2\rho},0,...,0)$, $\bvec{\hat{\psi}}_{rt} \equiv {}^t(0,...,0) $.
Especially, for $X_k$, $T_k$, $v_k$, and $Z_k$, the field amplitude $\rho$ is set to $\rho_{m,k}$ satisfying
\begin{equation}
F_k(\rho_{m,k})=0,
\label{rho_m_def}
\end{equation}
where $\rho_{m,k}$ is the renormalized spontaneous magnetization.
Note that in Eq.~(\ref{Def_Z}) the field renormalization factors $Z_k$ are defined as the momentum derivatives of $\Gamma_{1,k}^{(2)}$ with $\omega=p_x v_k$, and not $\omega=0$ as in Eq.~(\ref{Def_X}).
If the approximation Eq.~(\ref{Gamma_1}) is appropriate, $Z_k$ should be almost independent of the momentum and frequency at which the derivatives are evaluated.
However, in some cases, the contribution of other terms that do not appear in Eq.~(\ref{Gamma_1}) may not be negligible.
Thus, there is ambiguity in the choice of the momentum and frequency in the definition of $X_k$ and $Z_k$.
In Eq.~(\ref{Def_Z}), we have chosen the frequency such that, in the absence of the disorder, the RG equation for $Z_k$ is equivalent to that of the corresponding equilibrium model.
To derive the RG equations for $X_k$, $T_k$, $v_k$, $Z_k$ and $F_k(\rho)$, we require the exact flow equations for $\Gamma_{1,k}^{(1)}$ and $\Gamma_{1,k}^{(2)}$, which are obtained from the functional derivative of Eq.~(\ref{RG_Gamma_1}) (See Appendices \ref{sec:Appendix_RG_F} and \ref{sec:Appendix_RG_Z}).
The renormalized cumulant in Eq.~(\ref{Gamma_2}) is given by
\begin{equation}
\Delta^{\mu \nu}_k(\bvec{\psi}_{1},\bvec{\psi}_{2}) = \frac{1}{\Omega \mathcal{T}^2} \Gamma^{(11)}_{2; \hat{\psi}_1^{\mu} \hat{\psi}_2^{\nu} }(\mathbf{p},\omega) \bigr|_{\mathbf{p},\omega=0},
\label{Def_Delta}
\end{equation}
where the functional derivative is evaluated for a uniform field $ \bvec{\psi}_{1,rt} \equiv \bvec{\psi}_1, \: \bvec{\hat{\psi}}_{1,rt} \equiv \bvec{0} $ and  $ \bvec{\psi}_{2,rt} \equiv \bvec{\psi}_2, \: \bvec{\hat{\psi}}_{2,rt} \equiv \bvec{0} $.
The exact flow equation for $\Gamma_{2,k}^{(11)}$, which is obtained from the functional derivative of Eq.~(\ref{RG_Gamma_2}), leads to the RG equation for $\Delta^{\mu \nu}_k(\bvec{\psi}_{1},\bvec{\psi}_{2})$ (See Appendix \ref{sec:Appendix_RG_Delta}).

\subsection{Dimensionless parameters}
\label{sec:Dimensionless_parameters}

To obtain a fixed point, we express the RG equations in a scaled form by introducing renormalized dimensionless quantities.
For comparison between the equilibrium and nonequilibrium cases, we define the dimensionless quantities for both cases below.
In the following, the cutoff $\Lambda$ is set to unity.

\subsubsection{Equilibrium case}

For the equilibrium case ($v=0$), we follow the definition employed in Ref.~\onlinecite{Tarjus-08}.
Since the momentum $q$ is measured in units of $k$, we introduce the dimensionless momentum,
\begin{equation}
y = \frac{q^2}{k^2}.
\end{equation}
We define the dimensionless quantities, which are denoted with a tilde,
\begin{equation}
\rho = Z_k^{-1} k^{D-2} \tau_k^{-1} \tilde{\rho},
\label{rho_dimensionless_eq}
\end{equation}
\begin{equation}
F_k(\rho) = Z_{k} k^2 \tilde{F}_k(\tilde{\rho}),
\end{equation}
\begin{equation}
\Delta_k(\rho_1,\rho_2,z) = Z_{k} k^2 \tau_k^{-1} \tilde{\Delta}_k(\tilde{\rho}_1,\tilde{\rho}_2,z),
\label{Delta_dimensionless_eq}
\end{equation}
where $Z_k=Z_{\parallel,k}=Z_{\perp,k}$, and $\tau_k$ is chosen such that $\tilde{\Delta}_{k}$ attains a fixed point.
To define $\tau_k$, we introduce the renormalized disorder strength
\begin{equation}
\Delta_{m,k}=\Delta_{00,k}(\rho_{m,k},\rho_{m,k},z=1),
\label{Delta_m_def}
\end{equation}
where $\rho_{m,k}$ is given by Eq.~(\ref{rho_m_def}).
Then, $\tau_k$ is defined by
\begin{equation}
\tau_k=\frac{(Z_{k}/Z_{\Lambda}) k^2}{\Delta_{m,k}/\Delta_{m,\Lambda}}.
\label{tau_def_eq}
\end{equation}
From Eq.~(\ref{Delta_dimensionless_eq}), $\tilde{\Delta}_{00, k}(\rho_{m,k},\rho_{m,k},z=1)$ is constant along the RG flow.
We also introduce the running exponent associated with $\tau_k$ as
\begin{eqnarray}
\theta_k &=& k\partial_k \ln \tau_k \nonumber \\ 
&=& 2-\eta_{k}-k\partial_k \ln \Delta_{m,k},
\label{theta_def}
\end{eqnarray}
where $\eta_{k}$ is the running anomalous dimension
\begin{equation}
\eta_k = - k\partial_k \ln Z_k.
\label{eta_k}
\end{equation}

It is worth noting that $\tau_k$ can be considered as a renormalized temperature.
In the static formulation, the replicated Hamiltonian with temperature $\tau$ is given by
\begin{eqnarray}
H[\{ \bvec{\phi}_a \}] &=& \frac{1}{\tau} \sum_a \int d^D \bvec{r} \biggl[ \frac{1}{2} K |\nabla \bvec{\phi}_a|^2 + U(\rho_a) \biggr] \nonumber \\
&-& \frac{1}{2\tau^2} \sum_{a,b} \int d^D \bvec{r} h_0^2 \bvec{\phi}_a \cdot \bvec{\phi}_b. \nonumber
\end{eqnarray} 
The ratio of the kinetic term $|\nabla \bvec{\phi}_a|^2$ to the disorder term $h_0^2 \bvec{\phi}_a \cdot \bvec{\phi}_b$ is proportional to the temperature.
Since the kinetic term scales as $Z_k k^2$, Eq.~(\ref{tau_def_eq}) represents some kind of a renormalized temperature.
Note that $\tau_k$ should not be confused with $T_k$ in Eq.~(\ref{Gamma_1}), which is the strength of the renormalized thermal noise.
In equilibrium, $\tau_k \sim k^{\theta}$ with $\theta \simeq 2$, while $T_k$ is constant along the RG flow from the time-reversal symmetry Eq.~(\ref{time_reversal_trans_1}).
In the RG equations, $T_k$ and $\tau_k$ always appear in the combination $T_k \tau_k$ and the flow of this product controls the critical behavior of the system.

At a critical point, the ``connected'' and ``disconnected'' Green's functions behave as
\begin{equation}
G_{\rm c}(\bvec{r}-\bvec{r}') = \overline{ \frac{\delta \langle \bvec{\phi}(\bvec{r}) \rangle}{\delta h(\bvec{r}')} } \sim |\bvec{r}-\bvec{r}'|^{-(D-2+\eta)}, 
\end{equation}
\begin{equation}
G_{\rm d}(\bvec{r}-\bvec{r}') = \overline{\langle \bvec{\phi}(\bvec{r}) \rangle \langle \bvec{\phi}(\bvec{r}') \rangle} \sim |\bvec{r}-\bvec{r}'|^{-(D-4+\bar{\eta})},
\label{G_d_critical_eq}
\end{equation}
where $\eta$ is the fixed point value of $\eta_k$ and $\bar{\eta}$ is a new exponent.
Note that from the fluctuation-dissipation theorem $T G_{\rm c}(\bvec{r}-\bvec{r}')= \overline{ \langle \bvec{\phi}(\bvec{r})\bvec{\phi}(\bvec{r}') \rangle} - \overline{ \langle \bvec{\phi}(\bvec{r}) \rangle \langle \bvec{\phi}(\bvec{r}') \rangle } $.
Let us derive a relation between $\eta$, $\theta$, and $\bar{\eta}$.
The renormalized spontaneous magnetization $\rho_{m,k}$, which is defined in Eq.~(\ref{rho_m_def}), can be considered as the amplitude of the field averaged over a region $\mathcal{V}$ of linear dimension $k^{-1}$.
Therefore, at the critical point, it behaves as 
\begin{equation}
\rho_{m,k} \sim k^{2D} \int_{\mathcal{V}} d^D \bvec{r}  d^D \bvec{r}' G_{\mathrm{d}}(\bvec{r}-\bvec{r}') \sim k^{D-4+\bar{\eta}}, \nonumber
\end{equation}
where we have used Eq.~(\ref{G_d_critical_eq}).
On the other hand, from Eq.~(\ref{rho_dimensionless_eq}), $\rho_{m,k}$ scales as $k^{D-2+\eta-\theta}$ at the fixed point.
Thus, if we define 
\begin{equation}
\bar{\eta}_k = 2+\eta_k-\theta_k,
\end{equation}
and then, $\bar{\eta}$ is the fixed point value of $\bar{\eta}_k$.

The cutoff function $R_k(\mathbf{q})$ is written as
\begin{eqnarray}
R_k(\mathbf{q}) =  Z_{k} k^2 r(y),
\label{cutoff_function_eq}
\end{eqnarray}
where $y$ is the dimensionless momentum.
In the following calculations, we employ the ``optimized'' cutoff function,
\begin{eqnarray}
r(y) = (1-y) \Theta(1-y),
\label{optimized_cutoff_function}
\end{eqnarray}
where $\Theta(x)$ is a step function. \cite{Litim}

\subsubsection{Nonequilibrium case}

For the nonequilibrium case ($v \neq 0$), considering the anisotropy due to the driving, the transverse momentum $\mathbf{q}_{\perp}$ and longitudinal momentum $q_x$ are measured in units of $k$ and $Q_k$ (defined below), respectively.
Thus, we introduce the dimensionless momentum,
\begin{equation}
y_{\perp} = \frac{|\mathbf{q}_{\perp}|^2}{k^2}, \:\:\: y_{\parallel} = \frac{q_x^2}{Q_k^2}.
\end{equation}
We define the dimensionless quantities, which are denoted with a tilde,
\begin{equation}
\rho = Z_{\perp,k}^{-1} k^{D-3} Q_k \tau_k^{-1} \tilde{\rho},
\label{rho_dimensionless_neq}
\end{equation}
\begin{equation}
F_k(\rho) = Z_{\perp,k} k^2 \tilde{F}_k(\tilde{\rho}),
\end{equation}
\begin{equation}
\Delta_k(\rho_1,\rho_2,z) = Z_{\perp,k} k^2 \tau_k^{-1} \tilde{\Delta}_k(\tilde{\rho}_1,\tilde{\rho}_2,z).
\end{equation}
The temperature scaling exponent $\theta_k$ is defined by Eq.~(\ref{theta_def}) with
\begin{equation}
\eta_{\perp,k}=- k \partial_k \ln Z_{\perp,k}.
\label{Def_eta_perp}
\end{equation}
We also introduce the dimensionless velocity $\tilde{v}_k$ and dimensionless longitudinal elastic constant $\tilde{z}_{\parallel,k}$ as
\begin{equation}
\tilde{v}_k = \frac{X_k v_k Q_k}{Z_{\perp, k} k^2},
\end{equation}
\begin{equation}
\tilde{z}_{\parallel,k} = \frac{Z_{\parallel, k} Q_k^2}{Z_{\perp, k} k^2}.
\label{Def_z_parallel}
\end{equation}
$Q_k$ is chosen such that $\tilde{v}_k$ attains a fixed point.
Thus, we define
\begin{equation}
Q_k = \frac{(Z_{\perp, k}/Z_{\perp, \Lambda}) k^2}{X_k v_k/(X_{\Lambda} v_{\Lambda})}.
\end{equation}
The running exponent associated with $Q_k$ reads
\begin{eqnarray}
\zeta_k &=& k \partial_k \ln Q_k \nonumber \\
&=& 2-\eta_{\perp,k}- k \partial_k \ln (X_k v_k).
\label{Def_zeta}
\end{eqnarray}
Since $\tilde{v}_k$ is constant along the RG flow, we omit the subscript $k$ below.

At a critical point, the Green's functions for the transverse direction behave as
\begin{equation}
G_{\rm c}(\bvec{r}_{\perp}) \sim r_{\perp}^{-(D-2+\eta_{\perp})}, 
\end{equation}
\begin{equation}
G_{\rm d}(\bvec{r}_{\perp}) \sim r_{\perp}^{-(D-3+\bar{\eta}_{\perp})},
\label{G_d_critical_neq_perp}
\end{equation}
where $\eta_{\perp}$ is the fixed point value of $\eta_{\perp,k}$ and $\bar{\eta}_{\perp}$ is a new exponent.
For the longitudinal direction, they behave as
\begin{equation}
G_{\rm c}(r_{\parallel}) \sim r_{\parallel}^{-(D-2+\eta_{\perp})/\zeta}, 
\end{equation}
\begin{equation}
G_{\rm d}(r_{\parallel}) \sim r_{\parallel}^{-(D-3+\bar{\eta}_{\perp})/\zeta},
\label{G_d_critical_neq_para}
\end{equation}
where $\zeta$ is the fixed point value of $\zeta_k$.
Let us derive a relation between $\eta_{\perp}$, $\theta$, $\zeta$, and $\bar{\eta}_{\perp}$.
The renormalized spontaneous magnetization $\rho_{m,k}$ can be considered as the amplitude of the field averaged over a region $\mathcal{V}$ of linear dimension $k^{-1}$ for the perpendicular direction and $Q_k^{-1}$ for the parallel direction.
Therefore, at the critical point, it behaves as 
\begin{equation}
\rho_{m,k} \sim k^{2(D-1)}Q_k^2 \int_{\mathcal{V}} d^D \bvec{r}  d^D \bvec{r}' G_{\mathrm{d}}(\bvec{r}-\bvec{r}') \sim k^{D-3+\bar{\eta}_{\perp}}, \nonumber
\end{equation}
where we have used Eqs.~(\ref{G_d_critical_neq_perp}) and (\ref{G_d_critical_neq_para}).
On the other hand, from Eq.~(\ref{rho_dimensionless_neq}), $\rho_{m,k}$ scales as $k^{D-3+\zeta+\eta_{\perp}-\theta}$. 
Thus, if we define 
\begin{equation}
\bar{\eta}_{\perp,k} = \zeta_k+\eta_{\perp,k}-\theta_k,
\end{equation}
and then, $\bar{\eta}_{\perp}$ is the fixed point value of $\bar{\eta}_{\perp,k}$.

We assume that $R_k(\mathbf{q})=R_k(q_x^2, |\mathbf{q}_{\perp}|^2)$ is independent of the longitudinal momentum $q_x$,
\begin{eqnarray}
R_k(q_x^2, |\mathbf{q}_{\perp}|^2) = R_k(|\mathbf{q}_{\perp}|^2) =  Z_{\perp,k} k^2 r(y_{\perp}),
\label{cutoff_function}
\end{eqnarray}
where $r(y)$ is given by Eq.~(\ref{optimized_cutoff_function}).

\subsection{RG equations near the lower critical dimension}
\label{sec:RG_equation}

We show the RG equations at $D=4+\epsilon$ for the equilibrium case and at $D=3+\epsilon$ for the nonequilibrium case, in the first order of $\epsilon$.
In the case of the pure $O(N)$ model and near the lower critical dimension, the fixed point associated with the critical point is expected for a value of $\tilde{\rho}_m$ that diverges as $1/\epsilon$.
Therefore, one can organize a systematic expansion in powers of $1/\tilde{\rho}_m$.
The leading order of this expansion is considered below.

The transverse and longitudinal components for the one-replica propagator are denoted as $P^{(T)}(q)$ and $P^{(L)}(q)$, respectively, whose expressions are given in Appendix \ref{sec:Appendix_Green_function}.
If we assume that $\tilde{F}_k'(\tilde{\rho}_m)$ has a value of $\mathcal{O}(1)$ at the fixed point, $P^{(T)}(q) = \mathcal{O}(1)$ and $P^{(L)}(q) \sim \tilde{\rho}^{-1}$.
Therefore, retaining the leading order of the expansion of $1/\tilde{\rho}_m$ implies that the contribution from the longitudinal mode is ignored.

The RG flow of the one-replica force $F_k(\rho)$ is given by Eqs.~(\ref{RG_F}), (\ref{RG_F1}), and (\ref{RG_F2}) in Appendix \ref{sec:Appendix_RG_F}.
We introduce a scale parameter $l=-\ln (k/\Lambda)$, which moves from $0$ to $\infty$ as $k$ decreases from $\Lambda$ to $0$.
The RG equation for $\rho_m$ can be derived from $F_k'(\rho_m) \partial_l \rho_m + \partial_l F_k(\rho) |_{\rho=\rho_m} = 0$, where $\partial_l = - k \partial_k$.
In the leading order of $\rho_m^{-1}$, the RG equation for $\rho_m$ reads
\begin{eqnarray}
\partial_l \rho_m = - (N-1) \biggl[ \frac{T_k}{2}  L_{2,m}^{(T)} + \Delta_{m} I_{12,m}^{(T)}  \biggr],
\label{RG_rho_m}
\end{eqnarray}
where the integrals $L$ and $I$ are defined by Eqs.~(\ref{L_def}) and (\ref{I_def}) in Appendix \ref{sec:Appendix_Green_function}, respectively. 
We have used notations such as $I_{12,m}^{(T)}=I_{12}^{(T)}(\rho_m,\rho_m)$.
The renormalized disorder strength $\Delta_{m}$ is defined by Eq.~(\ref{Delta_m_def}).

We next consider the RG flows of $X_k$, $v_k$, $T_k$, and $Z_k$.
Since we are interested in the static properties of the critical point at zero temperature, it is sufficient to consider the flows of $Z_{\perp,k}$ and $X_k v_k$.
The RG equations for these are obtained from the momentum derivatives of $\partial_l \Gamma_1^{(2)}$, which are given by Eqs.~(\ref{RG_Gamma_1-2}), (\ref{gamma_1-2_b-1}), and (\ref{gamma_1-2_b-2}) in Appendix \ref{sec:Appendix_RG_Z}, and
\begin{eqnarray}
\partial_l Z_{\perp} = - \frac{\Delta_m}{2 \rho_m} \int_{\mathbf{q}} \partial_l R_k(\mathbf{q}) D_0(\mathbf{q})^{-2} \biggl[  \nonumber \\
4 M_0(\mathbf{q}) \biggl( Z_{\perp} + R_k'(\mathbf{q}) + \frac{2}{D-1} |\mathbf{q}_{\perp}|^2 R_k''(\mathbf{q}) \biggr) \nonumber \\
- \frac{4}{D-1} |\mathbf{q}_{\perp}|^2 \bigl( Z_{\perp} + R_k'(\mathbf{q}) \bigr)^2 \biggl( \frac{4 M_0(\mathbf{q})^2}{D_0(\mathbf{q})} - 1 \biggr)  \biggr] ,
\label{RG_Z}
\end{eqnarray}
\begin{eqnarray}
\partial_l (Xv)= - \frac{\Delta_m}{2 \rho_m} Xv \int_{\mathbf{q}} \partial_l R_k(\mathbf{q}) \frac{2M_0(\mathbf{q})}{D_0(\mathbf{q})^2},
\label{RG_v}
\end{eqnarray}
where $M(\mathbf{q})$ and $D(\mathbf{q})=D(\mathbf{q},\omega=0)$ are defined in Appendix \ref{sec:Appendix_Green_function}, and $R_k'(\mathbf{q}) = \partial_{|\mathbf{q}_{\perp}|^2} R_k(|\mathbf{q}_{\perp}|^2)$.
The running exponents $\eta_{\perp,k}$ and $\zeta_k$ can be calculated from these equations.
For the equilibrium case, the RG equations can be obtained by the replacements $Z_{\perp} \to Z$, $|\mathbf{q}_{\perp}|^2 \to q^2$, $(D-1) \to D$, and $R_k'(\mathbf{q}) \to \partial_{q^2} R_k(q^2)$.

Finally, let us consider the RG flow of the cumulant of the random field $\Delta^{\mu \nu}(\bvec{\psi}_1,\bvec{\psi}_2)$, which can be rewritten as Eq.~(\ref{Delta_expansion}).
The details of the derivation of the RG equation are given in Appendix \ref{sec:Appendix_RG_Delta}.
It is worth noting that in the leading order of $\rho_m^{-1}$, the RG equations for $\Delta_{00}$ and $\Delta_{21}$ do not contain $\Delta_{12}$, $\Delta_{11}$, or $\Delta_{22}$.
Since the flows of $Z_{\perp,k}$ and $X_k v_k$ depend on only $\Delta_{00}$, it is sufficient to consider the flow of $\Delta_{00}$ and $\Delta_{21}$.
The RG equations for these are given by
\begin{eqnarray}
\partial_{l} \Delta_{00}(\rho_1,\rho_2,z) = \frac{1}{2} A_{00}(\rho_1,\rho_2,z) T_k L_2^{(T)}(\rho_1) \nonumber \\
+ A_{00}(\rho_1,\rho_2,z) \Delta_{00}(\rho_1,\rho_1,1) I_{21}^{(T)}(\rho_1,\rho_1)  \nonumber \\
+ B_{00}(\rho_1,\rho_2,z) I_{21}^{(T)}(\rho_1,\rho_2)  \nonumber \\
+ C_{00}(\rho_1,\rho_2,z) J_{21}^{(T)}(\rho_1,\rho_2) 
+ \mathrm{perm}(\rho_1,\rho_2),
\label{RG_Delta00}
\end{eqnarray}
\begin{eqnarray}
\partial_{l} \Delta_{21}(\rho_1,\rho_2,z) = \frac{1}{2} A_{21}(\rho_1,\rho_2,z) T_k L_2^{(T)}(\rho_1) \nonumber \\
+ A_{21}(\rho_1,\rho_2,z) \Delta_{00}(\rho_1,\rho_1,1) I_{21}^{(T)}(\rho_1,\rho_1)  \nonumber \\
+ B_{21}(\rho_1,\rho_2,z) I_{21}^{(T)}(\rho_1,\rho_2)  \nonumber \\
+ C_{21}(\rho_1,\rho_2,z) J_{21}^{(T)}(\rho_1,\rho_2) 
+ \mathrm{perm}(\rho_1,\rho_2),
\label{RG_Delta21}
\end{eqnarray}
where the integral $J$ is defined by Eq.~(\ref{J_def}).
$A$, $B$, $C$ in Eqs.~(\ref{RG_Delta00}) and (\ref{RG_Delta21}) are given as follows:
\begin{eqnarray}
A_{00} = (2\rho_1)^{-1} \bigl[ (N-1) (2\rho_1 \partial_{\rho_1}- z \partial_z) \Delta_{00} \nonumber \\
+ (1-z^2) \partial_{z}^2 \Delta_{00} - 2 z \Delta_{21} \bigr], \nonumber
\end{eqnarray}
\begin{eqnarray}
B_{00} = (4\rho_1\rho_2)^{-1/2} \bigl[ (N-2+z^2) \Delta_{00} \partial_z \Delta_{00} + z \Delta_{00}^2 \nonumber \\
- z (1-z^2) ( \Delta_{21} \partial_z \Delta_{00} + \Delta_{00} \partial_z^2 \Delta_{00}) \nonumber \\
+ (1-z^2)^2 \Delta_{21} \partial_z^2 \Delta_{00} + (1+z^2) \Delta_{21} \Delta_{00} \bigr], \nonumber
\end{eqnarray}
\begin{eqnarray}
C_{00} = - 2 (4\rho_1\rho_2)^{-1/2} ( \Delta_{00} + z \Delta_{21} ) (1-z^2) \partial_z \Delta_{00}, \nonumber
\end{eqnarray}
\begin{eqnarray}
A_{21} = (2\rho_1)^{-1} \bigl[ (N-1) (2\rho_1 \partial_{\rho_1} \Delta_{21} - \Delta_{21} - z \partial_z \Delta_{21})  \nonumber \\
+ (1-z^2) \partial_{z}^2 \Delta_{21} - 4 z \partial_z \Delta_{21} - 2 \partial_z \Delta_{00} \bigr], \nonumber
\end{eqnarray}
\begin{eqnarray}
B_{21} = (4\rho_1\rho_2)^{-1/2} \bigl[ (N-2+z^2) \Delta_{00} \partial_z \Delta_{21} \nonumber \\
- z (1-z^2) ( \Delta_{21} \partial_z \Delta_{21} + \Delta_{00} \partial_z^2 \Delta_{21}) \nonumber \\ 
+ (1-z^2)^2 \Delta_{21} \partial_z^2 \Delta_{21} + 2 \Delta_{21} (z \Delta_{00} + z^2 \Delta_{21}) \nonumber \\
- 2 ( \partial_z \Delta_{00} + \Delta_{21} +2 z \partial_z \Delta_{21} ) \nonumber \\
\times \bigl\{ (1-z^2) \Delta_{21} - z \Delta_{00} \bigr\} \bigr], \nonumber
\end{eqnarray}
\begin{eqnarray}
C_{21} = (4\rho_1\rho_2)^{-1/2} \bigl[ (N-2+z^2) (\Delta_{21})^2 \nonumber \\ 
- 2 z (1-z^2) ( 2 \Delta_{21} \partial_z \Delta_{21} + \partial_z \Delta_{00} \partial_z \Delta_{21}) \nonumber \\ 
- 2(1-z^2) \Delta_{00} \partial_z \Delta_{21} + 2z^2 \Delta_{21} \partial_z \Delta_{00} \nonumber \\ 
+ z^2 (\partial_z \Delta_{00})^2 + (1-z^2)^2 (\partial_z \Delta_{21})^2 \nonumber \\ 
+ (\Delta_{00} + z \Delta_{21})(\Delta_{00} + 3 z \Delta_{21})  \bigr]. \nonumber
\end{eqnarray}
The zero-temperature case $T=T_k=0$ is considered below, and the first terms in Eqs.~(\ref{RG_Delta00}) and (\ref{RG_Delta21}) are omitted.

\subsubsection{Equilibrium case}

For the equilibrium case ($v=0$), the RG equations in terms of the dimensionless quantities introduced in Sec. \ref{sec:Dimensionless_parameters} can be derived.
The integrals in Eqs.~(\ref{RG_Delta00}) and (\ref{RG_Delta21}) at $\rho_1=\rho_2=\rho_m$ are calculated as
\begin{eqnarray}
I_{21}^{(T)}(\rho_m,\rho_m) = J_{21}^{(T)}(\rho_m,\rho_m) = Z_k^{-2} k^{D-4} \frac{8}{D} A_D,
\end{eqnarray}
where ${A_D}^{-1}=2^{D+1} \pi^{D/2} \Gamma(D/2)$.
We define $\delta_{00}(z)$ and $\delta_{21}(z)$ by
\begin{eqnarray}
\delta_{00}(z) &=& \frac{16}{D} A_D \: \frac{\tilde{\Delta}_{00}(\tilde{\rho}_m,\tilde{\rho}_m,z)}{2 \tilde{\rho}_m}, \nonumber \\
\delta_{21}(z) &=& \frac{16}{D} A_D \: \frac{\tilde{\Delta}_{21}(\tilde{\rho}_m,\tilde{\rho}_m,z)}{2 \tilde{\rho}_m}.
\end{eqnarray}
From Eqs.~(\ref{rho_dimensionless_eq}) and (\ref{RG_rho_m}), the RG equation for $\tilde{\rho}_m$ reads
\begin{equation}
\partial_l \tilde{\rho}_m = (D-4+\bar{\eta}_k) \tilde{\rho}_m - (N-1) \delta_{00}(1) \tilde{\rho}_m,
\label{RG_tilde_rho_m_eq}
\end{equation}
where $\bar{\eta}_k$ is determined from the condition that $2 \tilde{\rho}_m \delta_{00}(1)$ is constant along the RG flow.
From Eq.~(\ref{RG_Z}), the anomalous dimension $\eta_k$ is calculated as
\begin{equation}
\eta_k = \delta_{00}(1).
\end{equation}

From Eqs.~(\ref{RG_Delta00}) and (\ref{RG_Delta21}), the following relation holds,
\begin{equation}
\delta_{21}(z) = \partial_z \delta_{00}(z),
\label{delta21_delta00}
\end{equation}
which is clearly seen when observing that $\Delta_{00}(z)$ and $\Delta_{21}(z)$ can be written as
\begin{eqnarray}
\Delta_{00} = \frac{1}{\sqrt{4 \rho_1 \rho_2}} \partial_z V, \:\:\: \Delta_{21} = \frac{1}{\sqrt{4 \rho_1 \rho_2}} \partial_z^2 V,
\end{eqnarray}
where $V(\rho_1,\rho_2,z)$ is the cumulant of the random potential.
By introducing a potential $R(z)$ by $\delta_{00}(z)=\partial_z R(z)$ and $\delta_{21}(z)=\partial_z^2 R(z)$, the RG equation for $R(z)$ at $D=4+\epsilon$ is given by
\begin{eqnarray}
\partial_l R(z) = - \epsilon R(z) + 2 (N-2) R'(1) R(z) \nonumber \\
+ \frac{1}{2} (N-1) (R'(z)-2z R'(1)) R'(z) \nonumber \\
 + \frac{1}{2} (1-z^2) \bigl\{   - R'(z)^2 +2 ( R'(1) - z R'(z) ) R''(z) \nonumber \\
 + (1-z^2) R''(z)^2 \bigr\}.
\label{RG_R_eq}
\end{eqnarray}
This RG equation was also derived from the one-loop perturbative RG calculation. \cite{Fisher-85-1,Feldman-00}

\subsubsection{Nonequilibrium case}

We now derive the RG equations for the nonequilibrium case.
Note that the dimensionless longitudinal elastic constant $\tilde{z}_{\parallel,k}$, which is defined by Eq.~(\ref{Def_z_parallel}), scales as $\tilde{z}_{\parallel,k} \sim k^{2(\zeta-1)}$.
Thus, since $\zeta > 1$, $\tilde{z}_{\parallel,k}$ can be set to zero for large scales.
Furthermore, from Eqs.~(\ref{Def_zeta}), (\ref{RG_Z}), and (\ref{RG_v}), 
\begin{equation}
\zeta_k = 2.
\end{equation} 
The integrals in Eqs.~(\ref{RG_Delta00}) and (\ref{RG_Delta21}) at $\rho_1=\rho_2=\rho_m$ are calculated as
\begin{eqnarray}
I_{21}^{(T)}(\rho_m,\rho_m) &=&  Z_{\perp,k}^{-2} k^{D-3} \frac{2}{D-1} A_{D-1} \tilde{v}^{-1}, \nonumber \\
J_{21}^{(T)}(\rho_m,\rho_m) &=& 0.
\label{threshold_function_neq}
\end{eqnarray}
Note that the second integral $J_{21}^{(T)}$ vanishes in contrast to the equilibrium case.
This is independent of the choice for the cutoff function $R_k(\mathbf{q}_{\perp})$.

We define $\delta_{00}(z)$ and $\delta_{21}(z)$ by
\begin{eqnarray}
\delta_{00}(z) &=& \frac{4}{D-1} A_{D-1} \tilde{v}^{-1} \: \frac{\tilde{\Delta}_{00}(\tilde{\rho}_m,\tilde{\rho}_m,z)}{2 \tilde{\rho}_m}, \nonumber \\
\delta_{21}(z) &=& \frac{4}{D-1} A_{D-1} \tilde{v}^{-1} \: \frac{\tilde{\Delta}_{21}(\tilde{\rho}_m,\tilde{\rho}_m,z)}{2 \tilde{\rho}_m}.
\end{eqnarray}
The RG equation for $\tilde{\rho}_m$ reads
\begin{equation}
\partial_l \tilde{\rho}_m = (D-3+\bar{\eta}_{\perp, k}) \tilde{\rho}_m - (N-1) \delta_{00}(1) \tilde{\rho}_m,
\label{RG_tilde_rho_m}
\end{equation}
and the anomalous dimension for the transverse direction $\eta_{\perp, k}$ is written as
\begin{equation}
\eta_{\perp, k} = \delta_{00}(1).
\end{equation}
The RG equation for $\delta_{00}(z)$ and $\delta_{21}(z)$ at $D=3+\epsilon$ is given by
\begin{eqnarray}
\partial_l \delta_{00}(z) = -\epsilon \delta_{00}(z) + (N-2+z^2) \delta_{00}(z) \partial_z \delta_{00}(z) \nonumber \\
+ z \delta_{00}(z)^2 + (1+z^2) \delta_{00}(z) \delta_{21}(z) \nonumber \\
-z(1-z^2) \bigl\{ \delta_{21}(z) \partial_z \delta_{00}(z) + \delta_{00}(z) \partial_z^2 \delta_{00}(z) \bigr\} \nonumber \\
+ (1-z^2)^2 \delta_{21}(z) \partial_z^2 \delta_{00}(z) \nonumber \\
+ (N-3) \delta_{00}(1) \delta_{00}(z) - (N-1) z \delta_{00}(1) \partial_z \delta_{00}(z) \nonumber \\
- 2z \delta_{00}(1) \delta_{21}(z) + (1-z^2) \delta_{00}(1) \partial_z^2 \delta_{00}(z),
\label{RG_delta00}
\end{eqnarray}
\begin{eqnarray}
\partial_l \delta_{21}(z) = -\epsilon \delta_{21}(z) + (N-2+5z^2) \delta_{00}(z)  \partial_z \delta_{21}(z) \nonumber \\
 + 2z \delta_{00}(z) \bigl\{ \partial_z \delta_{00}(z) + 2 \delta_{21}(z) \bigr\} \nonumber \\
-2 (1-z^2) \delta_{21}(z) \partial_z \delta_{00}(z) - 2 (1-2z^2) \delta_{21}(z)^2 \nonumber \\
-z(1-z^2) \bigl\{ \delta_{00}(z) \partial_z^2 \delta_{21}(z) + 5 \delta_{21}(z) \partial_z \delta_{21}(z) \bigr\} \nonumber \\
+ (1-z^2)^2 \delta_{21}(z) \partial_z^2 \delta_{21}(z) \nonumber \\
- 2 \delta_{00}(1) \bigl\{ \partial_z \delta_{00}(z) + \delta_{21}(z) \bigr\} \nonumber \\
- (N+3) z \delta_{00}(1) \partial_z \delta_{21}(z) \nonumber \\
+ (1-z^2) \delta_{00}(1) \partial_z^2 \delta_{21}(z).
\label{RG_delta21}
\end{eqnarray}
Since the two integrals in Eq.~(\ref{threshold_function_neq}) have different values, the relation in Eq.~(\ref{delta21_delta00}) does not hold.

\section{Results}
\label{sec:Results}

In this section, we obtain a fixed point by solving the RG equations derived in the previous section.
We calculate the anomalous dimensions $\eta$ and $\bar{\eta}$ as functions of $N$ and compare them to those predicted from the dimensional reduction.

\subsection{Equilibrium case}

For the equilibrium case, one can find the detailed analysis in Refs.~\onlinecite{Feldman-02}, \onlinecite{Tarjus-08} and \onlinecite{Feldman-00}.
We briefly summarize these results in this section.

If the second derivative of $R(z)$ at $z=1$ is assumed finite, the derivatives of Eq.~(\ref{RG_R_eq}) show that a fixed function $R^*(z)$ exists for $N>N_{\mathrm{DR}}=18$.
For $N<N_{\mathrm{DR}}$, there is a range of initial condition $R_{k=\Lambda}'(1)$ for which the RG flow of $R_k''(1)$ diverges at a finite scale $k^{-1}=\xi_{\mathrm{L}}$, known as the Larkin length.
In this case, the fixed function behaves as $R^*(z) \sim (1-z)^{3/2}$ near $z=1$ and the dimensional reduction breaks down.

Let us consider the detailed behavior of this non-analytic fixed point.
For $N>N_{\mathrm{c}} \simeq 2.83$, a non-zero fixed point does not exist below $D=4$ ($\epsilon<0$), while above $D=4$ ($\epsilon>0$), an unstable fixed point exists, which corresponds to the transition between a LRO phase and a disordered phase.
The critical exponents for several values of $N$ are calculated as
\begin{eqnarray}
\eta(N=3) &=& 5.5 |\epsilon|, \nonumber \\
\eta(N=4) &=& 0.78 |\epsilon|, \nonumber \\
\eta(N=5) &=& 0.42 |\epsilon|.
\end{eqnarray}

For $N<N_{\mathrm{c}}$, a stable fixed point exists below $D=4$ ($\epsilon<0$).
This fixed point corresponds to a QLRO, which is known as the Bragg glass in the vortex lattices of superconductors \cite{Giamarchi-94}.
The critical exponent for an XY model is calculated as
\begin{eqnarray}
\eta(N=2) = \frac{\pi^2}{9} |\epsilon| \simeq 1.10 |\epsilon|.
\end{eqnarray}
Above $D=4$ ($\epsilon>0$), a non-zero fixed point does not exist.

For $N<N_{\mathrm{c}}$ and $D=4$, the RF$O(N)$M exhibits a ``logarithmic QLRO'', in which the disconnected Green's function exhibits logarithmic power-law decay $G_{\mathrm{d}}(\bvec{r}) \sim (\ln r)^{-\alpha(N)}$, where $\alpha(N)$ is independent of the strength of the disorder.
In particular, for $N=2$ (XY model), we have
\begin{eqnarray}
 G_{\mathrm{d}}(\bvec{r}) \propto
  \begin{cases}
    r^{-\eta}, \:\:(r < \xi_{\mathrm{L}}), & \\
    (\ln r)^{-\alpha}, \:\:(r \gg \xi_{\mathrm{L}}), &
  \end{cases}
\end{eqnarray}
where $\eta=h_0^2/(8 \pi^2 K^2)$ and $\alpha = \pi^2/9$. \cite{Feldman-00,Chitra}
The Larkin length is given by $\xi_{\mathrm{L}} = \exp (c \eta^{-1})$, where $c$ is a universal constant.
Such a logarithmic QLRO is also predicted in the two-dimensional random anisotropy XY model with a dipole interaction. \cite{Feldman-97}

\begin{figure}
 \centering
 \includegraphics[width=0.45\textwidth]{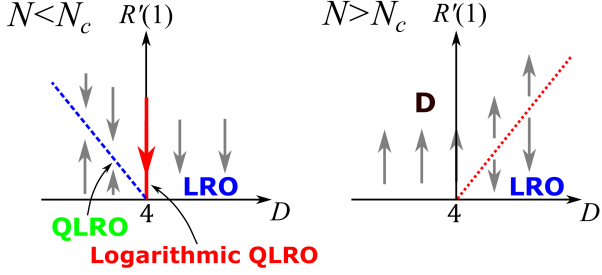}
 \caption{(Color online) Schematic picture of the RG flow of $R_k'(1)=\eta_k$ for the RF$O(N)$M.
 The horizontal axis represents the spatial dimension $D$.
 ``D'', ``LRO'', and ``QLRO'' denote the disordered, long-range order, and quasi-long-range order phases, respectively.}
 \label{fig:RG_flow_R}
\end{figure}

The schematic of the RG flow of $R'(1)=\eta_k$ is shown in Fig.~\ref{fig:RG_flow_R}. 
The blue (red) dashed (dotted) line represents stable (unstable) fixed points.
Note that for $N<N_{\mathrm{c}}$ we cannot obtain the fixed point corresponding to the second-order transition to a disordered phase in the leading order of $\epsilon$.
To obtain this fixed point, the beta function of $R(z)$ to the higher order of $\epsilon$ must be calculated. \cite{LeDoussal-06}

\subsection{Nonequilibrium case}

\subsubsection{Analytic fixed point}

Here we consider the nonequilibrium case Eqs.~(\ref{RG_delta00}) and (\ref{RG_delta21}).
First, the derivatives of $\delta_{00}(z)$ and $\delta_{21}(z)$ at $z=1$ are assumed finite.
By taking the derivative of Eqs.~(\ref{RG_delta00}) and (\ref{RG_delta21}),
\begin{eqnarray}
\partial_l \delta_{00}(1) = -\epsilon \delta_{00}(1) + (N-2) \delta_{00}(1)^2,
\label{RG_delta00(1)}
\end{eqnarray}
\begin{eqnarray}
\partial_l \delta_{00}'(1) = -\epsilon \delta_{00}'(1) + \delta_{00}(1)^2 + 2 \delta_{00}(1) \delta_{00}'(1) \nonumber \\
+ (N-1) \delta_{00}'(1)^2 + 4 \delta_{00}'(1) \delta_{21}(1),
\label{RG_delta00'(1)}
\end{eqnarray}
\begin{eqnarray}
\partial_l \delta_{21}(1) = -\epsilon \delta_{21}(1) + 2 \delta_{00}(1) \delta_{21}(1) + 2 \delta_{21}(1)^2.
\label{RG_delta21(1)}
\end{eqnarray}
From these equations, we have the following fixed points:
\begin{eqnarray}
\delta_{00}^*(1) &=& \frac{\epsilon}{N-2}, \nonumber \\
\delta_{00}'^*(1) &=& \frac{ (N-4) - \sqrt{(N-2)(N-10)}}{2(N-1)(N-2)} \epsilon, \nonumber \\
\delta_{21}^*(1) &=& 0,
\label{analytic_FP}
\end{eqnarray}
where we have chosen a stable solution for $\delta_{00}'^*(1)$ and $\delta_{21}^*(1)$.
In fact, Eq.~(\ref{RG_delta21(1)}) also has a fixed point $\delta_{21}^*(1)=(N-4)/2(N-2) \: \epsilon$, but it is unstable for $N>4$. 
The fixed point Eq.~(\ref{analytic_FP}) describes the second-order transition from a LRO phase to a disordered phase.  
Note that $\eta_{\perp} = \delta_{00}^*(1) = \epsilon/(N-2)$ is the same as the anomalous dimension for the $(2+\epsilon)$-dimensional pure $O(N)$ model.
This analytic fixed point exists for $N \geq 10$.
Thus,  $N_{\mathrm{DR}}=10$ is the field component number above which the dimensional reduction property is recovered.

\subsubsection{Non-analytic fixed point}

The non-analytic fixed points are now considered.
For the equilibrium case, $\delta_{00}$ and $\delta_{21}$ exhibit the following singularity near $z=1$,
\begin{eqnarray}
\delta_{00}^{(\mathrm{eq})}(z) &\sim& (1-z)^{1/2}, \nonumber \\
\delta_{21}^{(\mathrm{eq})}(z) = \partial_z \delta_{00}^{(\mathrm{eq})}(z) &\sim& -(1-z)^{-1/2}. \nonumber
\end{eqnarray}
Thus, we expand $\delta_{00}$ and $\delta_{21}$ around $z=1$,
\begin{eqnarray}
\delta_{00}(z) = a_0 + a_1 (1-z)^{1/2} + a_2 (1-z) + ... \nonumber \\
\delta_{21}(z) = b_{-1} (1-z)^{-1/2} + b_0 + b_1 (1-z)^{1/2} + ....
\label{delta_expand}
\end{eqnarray}
By substituting Eq.~(\ref{delta_expand}) into Eq.~(\ref{RG_delta21}), the right-hand side of Eq.~(\ref{RG_delta21}) yields a term proportional to $b_{-1}(1-z)^{-1}$, and we have
\begin{equation}
b_{-1} = 0.
\end{equation} 
Therefore, for the nonequilibrium case, $\delta_{21}(z)$ is finite at $z=1$, in contrast to the equilibrium case in which it diverges as $(1-z)^{-1/2}$.

By substituting Eq.~(\ref{delta_expand}) into Eq.~(\ref{RG_delta00}), we have the RG equation for $a_0=\delta_{00}(1)$,
\begin{equation}
\frac{d a_0}{dl} = -\epsilon a_0 + (N-2) a_0^2 - \frac{1}{2}(N-2)a_1^2 + 2 a_1 b_{-1},
\label{RG_a0}
\end{equation} 
where the third and fourth terms of the right-hand side are absent if $\delta_{00}(z)$ and $\delta_{21}(z)$ are analytic.
Since $b_{-1}=0$, the fixed point behaves as $\delta^*(z) \propto (N-2)^{-1}$ near $N=2$.
This fixed point is unstable and it corresponds to the second-order transition between the LRO phase and the disordered phase.
Note that the Bragg glass phase does not exist for $\epsilon<0$ and $N>2$.

\begin{figure}
 \centering
 \includegraphics[width=0.45\textwidth]{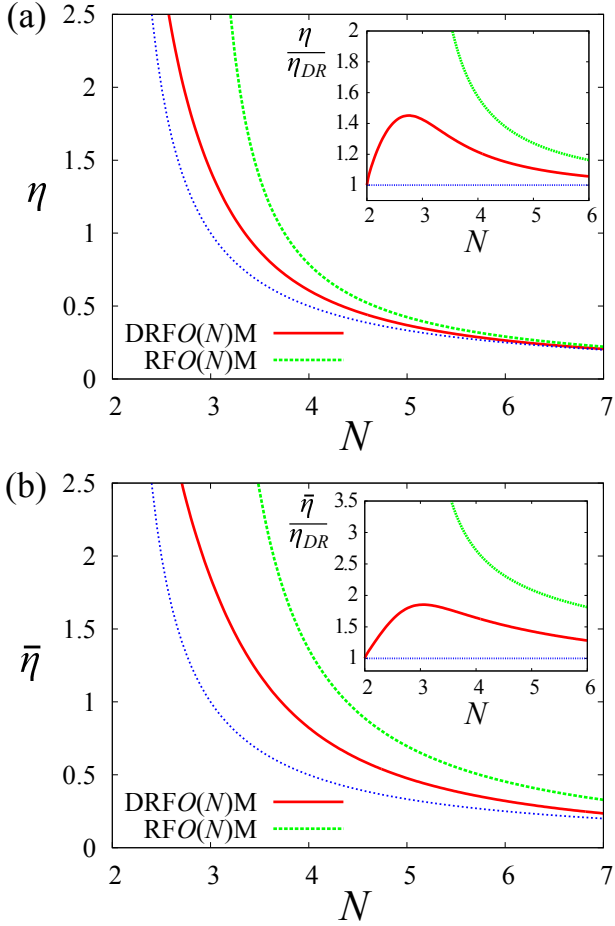}
 \caption{(Color online) Anomalous dimensions $\eta$ and $\bar{\eta}$ for the DRF$O(N)$M and RF$O(N)$M as functions of $N$.
 The red solid lines represent $\eta_{\perp}$ and $\bar{\eta}_{\perp}$ for the DRF$O(N)$M at $D=3+\epsilon$.
 The green dashed lines represent $\eta$ and $\bar{\eta}$ for the RF$O(N)$M at $D=4+\epsilon$.
 $\epsilon$ is set to unity.
 The blue dotted line represents the dimensional reduction value $\eta_{\mathrm{DR}}=(N-2)^{-1}$.
 The insets show $\eta/\eta_{\mathrm{DR}}$ and $\bar{\eta}/\eta_{\mathrm{DR}}$.}
 \label{fig:N-eta}
\end{figure}

\begin{figure}
 \centering
 \includegraphics[width=0.45\textwidth]{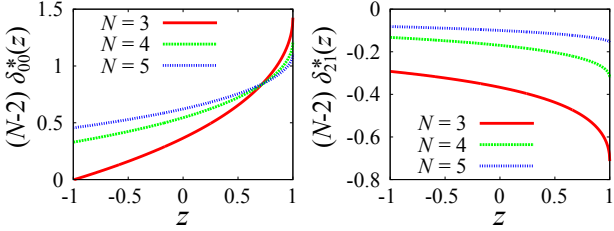}
 \caption{(Color online) Fixed functions $\delta_{00}^*(z)$ and $\delta_{21}^*(z)$ for $N=3$, $4$, and $5$.}
 \label{fig:delta*}
\end{figure}

The upper panel of Fig.~\ref{fig:N-eta} shows the anomalous dimension $\eta=\delta_{00}^*(1)$ for the nonequilibrium and equilibrium cases.
The red solid line represents $\eta_{\perp}$ for the nonequilibrium case at $D=3+\epsilon$ obtained from Eqs.~(\ref{RG_delta00}) and (\ref{RG_delta21}).
The green dashed line represents $\eta$ for the equilibrium case at $D=4+\epsilon$ obtained from Eq.~(\ref{RG_R_eq}).
The $\eta$ corresponding to the unstable fixed point is plotted.
Also shown is the anomalous dimension of the $(2+\epsilon)$-dimensional pure O(N) model, $\eta_{\mathrm{DR}}=(N-2)^{-1}$, which is indicated by the blue dotted line.
The inset displays the ratio of $\eta$ and $\eta_{\mathrm{DR}}$.
From Eqs.~(\ref{RG_tilde_rho_m_eq}) and (\ref{RG_tilde_rho_m}), the anomalous dimension for the disconnected Green's function $\bar{\eta}$ is given by
\begin{equation}
\bar{\eta} = - \epsilon + (N-1) \eta,
\end{equation}
which is shown in the lower panel of Fig.~\ref{fig:N-eta}.
For the equilibrium case, $\eta$ diverges as $N$ approaches to $N_{\mathrm{c}}=2.83$, while for the nonequilibrium case, $\eta_{\perp}$ diverges as $N$ approaches $2$.
The ratio $\eta_{\perp}/\eta_{\mathrm{DR}}$ exhibits non-monotonic behavior with decreasing $N$.
In particular,  $\lim_{N \to 2} \: \eta_{\perp}/\eta_{\mathrm{DR}} = 1$, which will be shown below.
Note that $\eta_{\perp}/\eta_{\mathrm{DR}}-1 \propto (N-2)$ near $N=2$, and
\begin{equation}
\lim_{N \to 2} (\eta_{\perp} - \eta_{\mathrm{DR}}) = \chi,
\end{equation}
where $\chi \simeq 1.65$ is a universal constant.
The fixed functions $\delta_{00}^*(z)$ and $\delta_{21}^*(z)$ of the DRF$O(N)$M are shown in Fig.~\ref{fig:delta*}.
Here, $\delta_{21}^*(z)$ significantly differs from $\partial_z \delta_{00}^*(z)$, in contrast to the equilibrium case where they are identical.
They exhibit non-analytic behavior in the form of $\sqrt{1-z}$ near $z=1$.
The numerical method to calculate the fixed function is explained in Appendix \ref{sec:Appendix_numerical}.

\subsubsection{Fixed line in the case that $N=2$ and $D=3$}

The 3D driven random field XY model (DRFXYM) is presented below.
The phase parameter $u$ is introduced and defined by $\bvec{\phi} = (\phi^1,\phi^2) = \sqrt{2 \rho} (\cos u , \sin u)$.
We denote the renormalized random force as $F^{\alpha}(\bvec{\phi})$, which satisfies 
\begin{equation}
\overline{ F^{\alpha}(\bvec{\phi}_1) F^{\beta}(\bvec{\phi}_2) } = \Delta^{\alpha \beta}(\bvec{\phi}_1,\bvec{\phi}_2).
\end{equation}
The tangential component of $F^{\alpha}(\bvec{\phi})$ is written as
\begin{equation}
F_{\perp}(u) = (2 \rho)^{-1/2} (-F^1(\bvec{\phi})\phi^2+F^2(\bvec{\phi})\phi^1),
\end{equation}
whose second cumulant is given by
\begin{eqnarray}
\Delta(u_1-u_2) &\equiv& \overline{ F_{\perp}(u_1) F_{\perp}(u_2) } \nonumber \\
&=& z \Delta_{00} - (1-z^2) \Delta_{21}.
\end{eqnarray}
Thus, we define a dimensionless cumulant $\delta(u)$ as
\begin{equation}
\delta(u) = z \delta_{00}(z) - (1-z^2) \delta_{21}(z),
\label{Def_delta_u}
\end{equation}
with $\cos u = z$.
Then, Eqs.~(\ref{RG_delta00}) and (\ref{RG_delta21}) can be reduced to the following simple equation:
\begin{eqnarray}
\partial_l \delta(u) = - \epsilon \delta(u) + \delta''(u) (\delta(0)-\delta(u)),
\label{RG_delta_XY}
\end{eqnarray}
which was also obtained by Ref.~\onlinecite{LeDoussal-98}.

In Ref.~\onlinecite{LeDoussal-98}, Eq.~(\ref{RG_delta_XY}) was derived from the driven random manifold model,
\begin{eqnarray}
\Gamma \bigl( \partial_t u + v \partial_x u \bigr) = K \nabla^2 u + F(\bvec{r};u) + \xi(\bvec{r},t),
\label{EM_SW}
\end{eqnarray}
which describes the dynamics of the displacement field $u(\bvec{r},t)$ of an elastic lattice moving in a random pinning potential.
The random force $F(\bvec{r};u)$ satisfies
\begin{equation}
\overline{ F(\bvec{r};u) F(\bvec{r'};u') } = \Delta(u-u') \delta(\bvec{r}-\bvec{r'}),
\end{equation}
where $\Delta(u)$ is a periodic function.
The driven random manifold model Eq.~(\ref{EM_SW}) is equivalent to the DRFXYM if the topological defects or vortices are ignored.
Therefore, Eq.~(\ref{EM_SW}) is an effective model that is valid for the weak disorder.

From Eq.~(\ref{RG_delta_XY}), at three dimensions ($\epsilon=0$), we find that the beta function of $\eta_{\perp,l}=\delta_l(0)$ vanishes identically even if $\delta(u)$ has a cusp at $u=0$.
This result can be also deduced from Eq.~(\ref{RG_a0}).
This suggests the existence of a QLRO characterized by {\it a line of fixed points} as in the 2D pure XY model.
The family of fixed points of Eq.~(\ref{RG_delta_XY}) at $\epsilon=0$ is given by
\begin{equation}
\delta^*(u) = C,
\end{equation}
or
\begin{eqnarray}
\begin{cases}
 \delta_{00}^*(z) = Cz, & \\
 \delta_{21}^*(z) = -C, &
\end{cases}
\label{FP-KT}
\end{eqnarray}
where $C$ is an arbitrary positive constant.
These fixed points are stable in the sense that $\delta_{00}(z)$ and $\delta_{21}(z)$ flow into Eq.~(\ref{FP-KT}) with an initial value dependent $C$ in the limit $l \to \infty$.
It is worth noting that the beta function of $\delta(0)$ does not vanish in the four-dimensional random field XY model. 
In this case, from Eq.~(\ref{RG_R_eq}), the RG equation is given by
\begin{eqnarray}
\partial_l \delta(u) = \delta''(u) (\delta(0)-\delta(u)) - \delta'(u)^2.
\label{RG_delta_XY_eq}
\end{eqnarray}
The last term on the right-hand side $\delta'(u)^2$ has a finite value at $u=0$ if $\delta(u)$ has a linear cusp.

Let us show that $\lim_{N \to 2} \: \eta_{\perp}/\eta_{\mathrm{DR}} = 1$ at $\epsilon=1$, as shown in Fig.~\ref{fig:N-eta}.
It is convenient to introduce $\tilde{\delta}_{00}(z) = (N-2) \delta_{00}(z)$ and $\tilde{\delta}_{21}(z) = (N-2) \delta_{21}(z)$.
The fixed functions $\tilde{\delta}_{00}^*(z)$ and $\tilde{\delta}_{21}^*(z)$ are stationary solutions of Eqs.~(\ref{RG_delta00}) and (\ref{RG_delta21}) with $\epsilon = (N-2)$.
Thus, in the limit $N \to 2$, $\tilde{\delta}_{00}^*(z)$ and $\tilde{\delta}_{21}^*(z)$ are given by Eq.~(\ref{FP-KT}).
The constant $C$ can be uniquely determined by considering a stationary solution of Eq.~(\ref{RG_a0}) with $\epsilon = (N-2)$.
Note that $a_1$ vanishes as $N \to 2$ because Eq.~(\ref{FP-KT}) does not have a cusp.
Thus, we have $a_0=C=1$, implying that $\lim_{N \to 2} \: \eta_{\perp}/\eta_{\mathrm{DR}} = 1$.

The existence of the fixed line implies that the 3D-DRFXYM may exhibit a Kosterlitz-Thouless (KT) transition. \cite{KT,Nelson}
However, there is a possibility that the beta function vanishes in the leading order but has a finite contribution in the higher order as in the case of the RF$O(N)$M at $D=4$ and $N=2.83$. \cite{Tissier-06-NPRG}
To verify the existence of the fixed line, {\it all higher order contributions} of the beta function must vanish.
A detailed investigation concerning this problem is a topic for future research.

\section{Summary}
\label{sec:Summary}

In this paper, we have studied the critical behavior of the DRF$O(N)$M at zero temperature.
From naive phenomenological arguments, we have introduced a dimensional reduction property, which states that the critical behavior of the $D$-dimensional DRF$O(N)$M at zero temperature is identical to that of the $(D-1)$-dimensional pure $O(N)$ model at finite temperature.
By employing the NPRG formalism, we found that the dimensional reduction breaks down for $2<N<10$ near three dimensions.
The deviation of the critical exponents from their dimensional reduction values is calculated in the first order of $\epsilon=D-3$.
It is worth noting that $\eta/\eta_{\mathrm{DR}}$ for the $(3+\epsilon)$-dimensional DRF$O(N)$M is smaller than that for the $(4+\epsilon)$-dimensional RF$O(N)$M.
This implies that the non-perturbative effect responsible for the failure of the dimensional reduction in driven disordered systems is weaker than that in equilibrium.

It is an intriguing question to ask whether the 2D driven random field Ising (DRFIM) model exhibits LRO.
The dimensional reduction implies that it does not have any ordered phase because it is identical to the one-dimensional pure Ising model. 
However, the lower critical dimension of the RFIM is two.
Since the driving reduces the lower critical dimension, it is expected that the 2D-DRFIM should exhibit LRO in the weak disorder and strong driving regime.
To clarify this question, the NPRG formalism developed in this study provides a promising approach.

The conventional dimensional reduction in equilibrium is a consequence of the supersymmetry (super-rotational invariance) of the stochastic field equation \cite{Parisi} and its breakdown can be understood as the spontaneous breaking of the supersymmetry. \cite{Tissier-12}
However, in Sec.~\ref{sec:DR}, the dimensional reduction for driven disordered systems is derived without introducing the supersymmetric formalism.
To obtain a clear insight into the underlying physics of its breakdown, we need to identify a hidden symmetry responsible for the dimensional reduction in driven disordered systems.

Finally, let us consider the role of temperature.
The thermal fluctuations lead to the local averaging of the disorder and thermally activated motion between different meta-stable states. 
At a finite temperature, the diffusion-like terms, which are proportional to $\tilde{T}_k \equiv T_k \tau_k \tilde{z}_{\parallel,k}^{-1/2}$, appear in the RG equation for the renormalized cumulant and they smooth out the cusp.
In equilibrium, the temperature is irrelevant near the fixed point controlling the critical behavior, $\tilde{T}_k = k^{\theta}$ with $\theta = 2 + \mathcal{O}(\epsilon)$.
However, in nonequilibrium cases, there is a possibility that a fixed point with non-zero temperature $\tilde{T}^* \neq 0$ may occur due to the additional terms, which result from the violation of the fluctuation-dissipation theorem, in the RG equation for $T_k$. 
For example, in the case of the DRFXYM, the RG flow of $\tilde{T}_k$ is given by
\begin{equation}
\frac{d \tilde{T}_k}{dl} = \bigl[ -1 - \epsilon - \delta''(0) \bigr] \tilde{T}_k,
\end{equation} 
where $\delta(u)$ is defined by Eq.~(\ref{Def_delta_u}) with $\cos u = z$. \cite{LeDoussal-98,Balents-98} 
Since $-\delta''(0)$ diverges as $\tilde{T}_k$ decreases to zero, the competition between the first two terms and the last term in the brackets leads to a non-zero fixed point.
The critical behavior of the DRF$O(N)$M at finite temperature could be also controlled by such non-zero temperature fixed points.
This implies that an infinitesimal thermal noise qualitatively changes the large-scale structure of the system.
A detailed investigation of finite temperature cases will be conducted in future studies.

\begin{acknowledgments}
We are grateful to Gilles Tarjus and Matthieu Tissier for their helpful discussions and remarks.
The present study was supported by JSPS KAKENHI Grant No. 15J01614, a Grant-in-Aid for JSPS Fellows.
\end{acknowledgments}

\appendix

\section{Field theoretical formalism for the driven disordered system}
\label{sec:Appendix_action}

In this appendix, we derive the disorder averaged action Eqs.~(\ref{S_1}) and (\ref{S_2}) from the Langevin equation (\ref{EM}).
We introduce $n$ replicas of the system, $\{ \bvec{\phi}_a \}_{a=1}^n$, with the same disorder.
The dynamics are given by
\begin{equation}
\Gamma (\partial_t + v \partial_x) \bvec{\phi}_a = K \nabla^2 \bvec{\phi}_a - U'(\rho_a) \bvec{\phi}_a + \bvec{h} + \bvec{\xi}_a,
\label{Appendix_EM}
\end{equation}
where the thermal noise satisfies
\begin{eqnarray}
\langle \xi_a^{\alpha}(\bvec{r},t) \xi_b^{\beta}(\bvec{r}',t') \rangle = 2\Gamma T \delta^{\alpha \beta} \delta_{ab} \delta(\bvec{r}-\bvec{r}')\delta(t-t').
\end{eqnarray}
The average of a function of the field $A[\{\bvec{\phi}_a\}]$ over the thermal noise is written as
\begin{eqnarray}
\langle A[\{\bvec{\phi}_a\}] \rangle = \int \mathcal{D} \xi P[\xi] \int \prod_a \mathcal{D} \bvec{\phi}_a \nonumber \\
\times \delta(\bvec{\phi}_a-\bvec{\phi}_a[\xi]) A[\{\bvec{\phi}_a\}],
\end{eqnarray}
where $\bvec{\phi}_a[\xi]$ is the solution of Eq.~(\ref{Appendix_EM}) for a realization of the noise $\xi_a$.
This average can be calculated as
\begin{eqnarray}
\langle A[\{\bvec{\phi}_a\}] \rangle = \int \mathcal{D} \xi P[\xi] \int \prod_a \mathcal{D} \bvec{\phi}_a \mathcal{J}[\{\bvec{\phi}_a\}]  \nonumber \\
\times A[\{\bvec{\phi}_a\}]  \delta \bigl[ \: \Gamma (\partial_t + v \partial_x) \bvec{\phi}_a - K \nabla^2 \bvec{\phi}_a \nonumber \\
 + U'(\rho_a) \bvec{\phi}_a - \bvec{h} - \bvec{\xi}_a \: \bigr] \nonumber \\
= \int \mathcal{D} \xi P[\xi] \int \prod_a \mathcal{D} \bvec{\phi}_a \mathcal{D} \bvec{\hat{\phi}}_a \mathcal{J}[\{\bvec{\phi}_a\}] A[\{\bvec{\phi}_a\}]  \nonumber \\
\times \exp \biggl[ - \sum_a \int_{rt} i \bvec{\hat{\phi}}_a \cdot \bigl\{ \Gamma (\partial_t + v \partial_x) \bvec{\phi}_a \nonumber \\
- K \nabla^2 \bvec{\phi}_a + U'(\rho_a) \bvec{\phi}_a - \bvec{h} - \bvec{\xi}_a \bigr\} \biggr] \nonumber \\
= \int \prod_a \mathcal{D} \bvec{\phi}_a \mathcal{D} \bvec{\hat{\phi}}_a \mathcal{J}[\{\bvec{\phi}_a\}] A[\{\bvec{\phi}_a\}] \exp \biggl[ \nonumber \\
 - \sum_a \int_{rt} i \bvec{\hat{\phi}}_a \cdot \bigl\{ \Gamma (\partial_t \bvec{\phi}_a + v \partial_x \bvec{\phi}_a - T i \bvec{\hat{\phi}}_a) \nonumber \\
- K \nabla^2 \bvec{\phi}_a  + U'(\rho_a) \bvec{\phi}_a - \bvec{h}  \bigr\} \biggr],
\end{eqnarray}
where $\mathcal{J}[\{\bvec{\phi}_a\}]$ is the Jacobian, which is set to unity. \cite{Canet-11}
We next take the average over the disorder $\bvec{h}$,
\begin{eqnarray}
\overline{\langle A[\{\bvec{\phi}_a\}] \rangle} = \int \prod_a \mathcal{D} \Phi_a  A[\{\bvec{\phi}_a\}] \exp \bigl( -S[\{ \Phi_a \}] \bigr),
\end{eqnarray}
where the disorder averaged action $S[\{ \Phi_a \}]$ is given by Eqs.~(\ref{S_1}) and (\ref{S_2}).
For simplicity , we have omitted $i$ in $i \bvec{\hat{\phi}}_a$.

\section{Exact flow equations for $\Gamma_{p,k}$}
\label{sec:Appendix_exact_flow_equations}

In this appendix, we derive the exact flow equations for $\Gamma_{p,k}$, Eqs.~(\ref{RG_Gamma_1}) and (\ref{RG_Gamma_2}).
To calculate the inverse of $\Gamma_k^{(2)} + \hat{\mathbf{R}}_k$ with respect to the replica indices, we rewrite it as
\begin{eqnarray}
\bigl( \Gamma_k^{(2)} + \hat{\mathbf{R}}_k \bigr)_{ab} = \mathrm{P}_k[\Psi_a]^{-1} \delta_{ab} - A[\Psi_a] \delta_{ab} \nonumber \\
- B[\Psi_a,\Psi_b],
\label{Appendix_Gamma(2)_R}
\end{eqnarray}
where $\mathrm{P}_k[\Psi]$ is the one-replica propagator defined by Eq.~(\ref{Green_function}).
$A[\Psi_a]$ and $B[\Psi_a,\Psi_b]$ can be expanded in an increasing number of free replica sums,
\begin{eqnarray}
A[\Psi_a] = \sum_c A^{[1]}[\Psi_a|\Psi_c] \nonumber \\
+ \frac{1}{2} \sum_{c,d} A^{[2]}[\Psi_a|\Psi_c,\Psi_d] +...,
\label{Appendix_A_expansion}
\end{eqnarray}
\begin{eqnarray}
B[\Psi_a,\Psi_b] = B^{[0]}[\Psi_a,\Psi_b] + \sum_c B^{[1]}[\Psi_a,\Psi_b|\Psi_c] \nonumber \\
+ \frac{1}{2} \sum_{c,d} B^{[2]}[\Psi_a,\Psi_b|\Psi_c,\Psi_d] + ...,
\label{Appendix_B_expansion}
\end{eqnarray}
where the vertical bar in each term $A^{[p]}[\Psi_a|\Psi_{c_1},...,\Psi_{c_p}]$ is introduced to distinguish between the ``explicit'' index $a$ and the dummy indices $c_1,...,c_p$, which run from $1$ to $n$ as the summation is taken.
In the following, we use simplified notations such as,
\begin{eqnarray}
\Gamma_3^{(200)}[\Psi_1,\Psi_2,\Psi_3] = \frac{\delta^2 \Gamma_3[\Psi_1,\Psi_2,\Psi_3]}{\delta \Psi_1 \delta \Psi_1}, \nonumber \\
\Gamma_3^{(110)}[\Psi_1,\Psi_2,\Psi_3] = \frac{\delta^2 \Gamma_3[\Psi_1,\Psi_2,\Psi_3]}{\delta \Psi_1 \delta \Psi_2}.
\end{eqnarray}
From Eq.~(\ref{Gamma_expansion}), $A^{[p]}$ and $B^{[p]}$ are written as
\begin{eqnarray}
A^{[1]}[\Psi_a|\Psi_c] &=& \Gamma_2^{(20)}[\Psi_a,\Psi_c], \nonumber \\
A^{[2]}[\Psi_a|\Psi_c,\Psi_d] &=& - \Gamma_3^{(200)}[\Psi_a,\Psi_c,\Psi_d], \nonumber \\
...,
\label{Appendix_A_Gamma}
\end{eqnarray}
and
\begin{eqnarray}
B^{[0]}[\Psi_a,\Psi_b] &=& \Gamma_2^{(11)}[\Psi_a,\Psi_b], \nonumber \\
B^{[1]}[\Psi_a,\Psi_b|\Psi_c] &=& - \Gamma_3^{(110)}[\Psi_a,\Psi_b,\Psi_c], \nonumber \\
B^{[2]}[\Psi_a,\Psi_b|\Psi_c,\Psi_d] &=& \Gamma_4^{(1100)}[\Psi_a,\Psi_b,\Psi_c,\Psi_d], \nonumber \\
....
\label{Appendix_B_Gamma}
\end{eqnarray}
The inverse of Eq.~(\ref{Appendix_Gamma(2)_R}) reads
\begin{eqnarray}
\bigl( \Gamma_k^{(2)} + \hat{\mathbf{R}}_k \bigr)^{-1}_{ab} = \mathrm{P}_k[\Psi_a] \delta_{ab} \nonumber \\
 + \mathrm{P}_k[\Psi_a]\bigl( A[\Psi_a] \delta_{ab}+B[\Psi_a,\Psi_b]\bigr)\mathrm{P}_k[\Psi_b] \nonumber \\
+ \mathrm{P}_k[\Psi_a]\bigl( A[\Psi_a] \delta_{ac}+B[\Psi_a,\Psi_c] \bigr)\mathrm{P}_k[\Psi_c] \nonumber \\
\times \bigl( A[\Psi_c] \delta_{cb}+B[\Psi_c,\Psi_b] \bigr)\mathrm{P}_k[\Psi_b] +....
\end{eqnarray}
By substituting Eqs.~(\ref{Appendix_A_expansion}) and (\ref{Appendix_B_expansion}) into the above equation, we have
\begin{eqnarray}
\sum_a \bigl( \Gamma_k^{(2)} + \hat{\mathbf{R}}_k \bigr)^{-1}_{aa} = \sum_a Q_1[\Psi_a] \nonumber \\
+ \frac{1}{2} \sum_{a,b} Q_2[\Psi_a,\Psi_b] + \frac{1}{3!} \sum_{a,b,c} Q_3[\Psi_a,\Psi_b,\Psi_c] +...,
\end{eqnarray}
where $Q_1$ and $Q_2$ are given by
\begin{eqnarray}
Q_1[\Psi_a] = \mathrm{P}_k[\Psi_a] + \mathrm{P}_k[\Psi_a] B^{[0]}[\Psi_a,\Psi_a] \mathrm{P}_k[\Psi_a],
\label{Appendix_Q_1}
\end{eqnarray}
\begin{eqnarray}
Q_2[\Psi_a,\Psi_b] = \mathrm{P}_k[\Psi_a] \bigl\{ A^{[1]}[\Psi_a|\Psi_b]  \nonumber \\
+ B^{[1]}[\Psi_a,\Psi_a|\Psi_b] \nonumber \\
+ B^{[0]}[\Psi_a,\Psi_b] \mathrm{P}_k[\Psi_b] B^{[0]}[\Psi_b,\Psi_a]  \nonumber \\
+ A^{[1]}[\Psi_a|\Psi_b] \mathrm{P}_k[\Psi_a] B^{[0]}[\Psi_a,\Psi_a] \nonumber \\
+ B^{[0]}[\Psi_a,\Psi_a] \mathrm{P}_k[\Psi_a] A^{[1]}[\Psi_a|\Psi_b]  \bigr\} \mathrm{P}_k[\Psi_a] \nonumber \\
+ \mathrm{perm}(\Psi_a,\Psi_b).
\label{Appendix_Q_2}
\end{eqnarray}
From Eqs.~(\ref{Appendix_A_Gamma}), (\ref{Appendix_B_Gamma}), (\ref{Appendix_Q_1}), and (\ref{Appendix_Q_2}), we obtain the exact flow equations for $\Gamma_1$ and $\Gamma_2$, Eqs.~(\ref{RG_Gamma_1}) and (\ref{RG_Gamma_2}).

\section{RG equation for $F_k$}
\label{sec:Appendix_RG_F}

\begin{figure}
 \centering
 \includegraphics[width=0.4\textwidth]{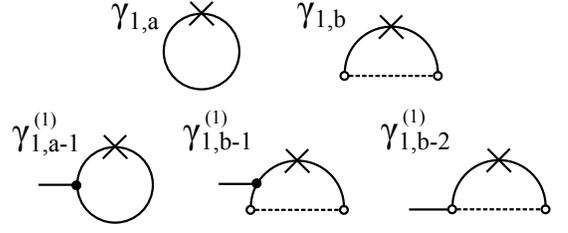}
 \caption{Graphical representations for the flow equations of $\Gamma_{1}$ and $\Gamma_{1}^{(1)}$.}
 \label{fig:gamma_1-1}
\end{figure}

In this Appendix, the RG equation for $F_k(\rho)$ is derived, which is given by Eq.~(\ref{Def_F}).
To do this, the exact flow equation for $\Gamma_1^{(1)}$ is required.
It is convenient to introduce a graphical representation.
The flow equation for $\Gamma_1$ Eq.~(\ref{RG_Gamma_1}) is rewritten as
\begin{equation}
\partial_l \Gamma_{1} = \frac{1}{2} \bigl[ \gamma_{1,a} + \gamma_{1,b} \bigr],
\label{RG_Gamma_1_graph}
\end{equation}
where $\gamma_{1,a}$ and $\gamma_{1,b}$ are given in Fig. \ref{fig:gamma_1-1}.
A scale parameter $l=-\ln (k/\Lambda)$ is introduced.
The flow equation for $\Gamma_{1;\hat{\psi}^1}^{(1)}=\delta \Gamma_1/\delta \hat{\psi}^1$ is then written as
\begin{equation}
\partial_l \Gamma_{1;\hat{\psi}^1}^{(1)} = \frac{1}{2} \bigl[ -\gamma_{1,a-1}^{(1)} - 2 \gamma_{1,b-1}^{(1)} + 2 \gamma_{1,b-2}^{(1)} \bigr].
\label{RG_Gamma_1-1}
\end{equation}
where $\gamma_{1,a-1}^{(1)}$, $\gamma_{1,b-1}^{(1)}$, and $\gamma_{1,b-2}^{(1)}$ are also given in Fig. \ref{fig:gamma_1-1}.
The rule for the graphical representation is as follows:
\begin{enumerate}
 \item An inner line denotes the propagator $\mathrm{P}[\Psi]$.
 \item A filled circle represents a vertex obtained from a derivative of the one-replica action $\Gamma_1^{(p)}[\Psi]$.
 \item Two open dots linked by a dashed line represent vertex obtained from a derivative of the two-replica action $\Gamma_2^{(p_1 p_2)}[\Psi_1,\Psi_2]$.
 \item A cross symbol denotes $\partial_l \bvec{\mathrm{R}}_k(\mathbf{q})$.
\end{enumerate}
For example, $\gamma_{1,b-1}^{(1)}$ is written as
\begin{eqnarray}
\gamma_{1,b-1}^{(1)} = \mathrm{Tr} \bigl[ \partial_l \bvec{\mathrm{R}}_k(\mathbf{q}) \mathrm{P}[\Psi]  \Gamma_2^{(11)}[\Psi,\Psi] \nonumber \\
\times \mathrm{P}[\Psi] \Gamma_{1;\hat{\psi}^1}^{(3)}[\Psi] \mathrm{P}[\Psi] \bigr],
\end{eqnarray}
where $\Gamma_{1;\hat{\psi}^1}^{(3)}=\delta \Gamma_{1}^{(2)}/\delta \hat{\psi}^1$.
All functional derivatives are evaluated for a uniform field $ \bvec{\psi}_{rt} \equiv {}^t(\sqrt{2\rho},0,...,0)$, $\bvec{\hat{\psi}}_{rt} \equiv {}^t(0,...,0) $.

The following notation is introduced:
\begin{eqnarray}
\Gamma_{1; \psi^2 \psi^2 \hat{\psi}^1}^{(3)}(q_1,q_2,q_3) = \frac{\delta^2 \Gamma_1[\Psi]}{\delta \psi^2(q_1) \delta \psi^2(q_2) \delta \hat{\psi}^1(q_3)}, \nonumber \\
\Gamma_{2; \psi_1^2 \hat{\psi}_1^2 \hat{\psi}_2^1}^{(21)}(q_1,q_2,q_3) = \frac{\delta^2 \Gamma_2[\Psi_1,\Psi_2]}{\delta \psi_1^2(q_1) \delta \hat{\psi}_1^2(q_2) \delta \hat{\psi}_2^1(q_3)}. \nonumber
\end{eqnarray}
From Eqs.~(\ref{Gamma_1}) and (\ref{Gamma_2}), they are calculated as
\begin{eqnarray}
\Gamma_{1;\psi^1 \psi^1 \hat{\psi}^1}^{(3)}(q_1,q_2,q_3) = - \sqrt{2\rho} (3F'(\rho)+2\rho F''(\rho)) \nonumber \\ 
\times (2\pi)^{D+1}\delta(\mathbf{q}_1+\mathbf{q}_2+\mathbf{q}_2)\delta(\omega_1+\omega_2+\omega_2), \nonumber 
\end{eqnarray}
\begin{eqnarray}
\Gamma_{1;\psi^{\nu} \psi^{\nu} \hat{\psi}^1}^{(3)}(q_1,q_2,q_3) = - \sqrt{2\rho} F'(\rho) \nonumber \\
\times (2\pi)^{D+1}\delta(\mathbf{q}_1+\mathbf{q}_2+\mathbf{q}_2)\delta(\omega_1+\omega_2+\omega_2),  \nonumber \\
(\nu=2,...,N), \nonumber
\end{eqnarray}
\begin{eqnarray}
\Gamma_{2; \hat{\psi}_1^1 \hat{\psi}_2^1}^{(11)}(q_1,q_2)\bigr|_{\Psi_1=\Psi_2=\Psi} = \Delta_L(\rho) \nonumber \\
\times (2\pi)^{D+2}\delta(\mathbf{q}_1+\mathbf{q}_2)  \delta(\omega_1) \delta(\omega_2), \nonumber
\end{eqnarray}
\begin{eqnarray}
\Gamma_{2; \hat{\psi}_1^{\nu} \hat{\psi}_2^{\nu}}^{(11)}(q_1,q_2)\bigr|_{\Psi_1=\Psi_2=\Psi}  = \Delta_T(\rho) \nonumber \\
\times (2\pi)^{D+2}\delta(\mathbf{q}_1+\mathbf{q}_2)  \delta(\omega_1) \delta(\omega_2), \nonumber \\
(\nu=2,...,N), \nonumber
\end{eqnarray}
\begin{eqnarray}
\Gamma_{2; \psi_1^1 \hat{\psi}_1^1 \hat{\psi}_2^1}^{(21)}(q_1,q_2,q_3)\bigr|_{\Psi_1=\Psi_2=\Psi} = \frac{1}{2} \sqrt{2 \rho} \Delta_L'(\rho) \nonumber \\
\times (2\pi)^{D+2}\delta(\mathbf{q}_1+\mathbf{q}_2+\mathbf{q}_3) \delta(\omega_1+\omega_2) \delta(\omega_3), \nonumber 
\end{eqnarray}
\begin{eqnarray}
\Gamma_{2; \psi_1^{\nu} \hat{\psi}_1^1 \hat{\psi}_2^{\nu}}^{(21)}(q_1,q_2,q_3)\bigr|_{\Psi_1=\Psi_2=\Psi} =\frac{1}{\sqrt{2 \rho}} \bigl( \Delta_{21}(\rho)+\Delta_{11}(\rho) \bigr) \nonumber \\
\times (2\pi)^{D+2}\delta(\mathbf{q}_1+\mathbf{q}_2+\mathbf{q}_3) \delta(\omega_1+\omega_2) \delta(\omega_3), \nonumber \\
(\nu=2,...,N), \nonumber
\end{eqnarray}
where we have used the notations $\Delta_{...}(\rho)=\Delta_{...}(\rho,\rho,z=1)$, $\Delta_T(\rho) = \Delta^{22}(\bvec{\psi},\bvec{\psi}) = \Delta_{00}(\rho)$, and $\Delta_L(\rho) = \Delta^{11}(\bvec{\psi},\bvec{\psi})=\Delta_{00}(\rho)+\Delta_{12}(\rho)+\Delta_{21}(\rho)+\Delta_{11}(\rho)+\Delta_{22}(\rho)$.
From these expressions, we obtain
\begin{eqnarray}
\gamma_{1,a-1}^{(1)} = \sqrt{2\rho} T \Bigl[ (N-1) F_k'(\rho)L_2^{(T)}(\rho) \nonumber \\
 + (3F'(\rho)+2\rho F''(\rho)) L_2^{(L)}(\rho)  \Bigr], \nonumber
\end{eqnarray}
\begin{eqnarray}
\gamma_{1,b-1}^{(1)} = \sqrt{2\rho} \Bigl[ (N-1) F'(\rho) \Delta_T(\rho) I_{12}^{(T)}(\rho) \nonumber \\
 + (3F'(\rho)+2\rho F''(\rho))  \Delta_L(\rho) I_{12}^{(L)}(\rho)  \Bigr], \nonumber
\end{eqnarray}
\begin{eqnarray}
\gamma_{1,b-2}^{(1)} = - \sqrt{2\rho} \biggl[ (N-1) \frac{1}{2 \rho} \bigl( \Delta_{21}(\rho)+\Delta_{11}(\rho) \bigr) J_{11}^{(T)}(\rho) \nonumber \\
+ \frac{1}{2}  \Delta_L'(\rho) J_{11}^{(L)}(\rho) \biggr], \nonumber
\end{eqnarray}
where we have already calculated the $\omega$-integral.
The functions $L$, $I$, and $J$ are defined by Eqs.~(\ref{L_def}), (\ref{I_def}), and (\ref{J_def}) in Appendix \ref{sec:Appendix_Green_function}, respectively, and simplified notations such as $I_{nn'}^{(T)}(\rho)=I_{nn'}^{(T)}(\rho,\rho)$ are used.
From Eqs.~(\ref{Def_F}) and (\ref{RG_Gamma_1-1}), we have the RG equation for $F(\rho)$,
\begin{eqnarray}
\partial_l F(\rho) = \partial_l F^{(1)}(\rho) + \partial_l F^{(2)}(\rho),
\label{RG_F}
\end{eqnarray}
where $\partial_l F^{(1)}(\rho)$ and $\partial_l F^{(2)}(\rho)$ are the contributions from the one and two-replica parts, respectively,
\begin{eqnarray}
\partial_l F^{(1)}(\rho) = \frac{1}{2} T \Bigl[ (N-1) F'(\rho) L_2^{(T)}(\rho) \nonumber \\
+ (3 F'(\rho) + 2 \rho F''(\rho)) L_2^{(L)}(\rho) \Bigr],
\label{RG_F1}
\end{eqnarray}
\begin{eqnarray}
\partial_l F^{(2)}(\rho) = (N-1) F'(\rho) \Delta_T(\rho) I_{12}^{(T)}(\rho) \nonumber \\
+ (3 F'(\rho) + 2 \rho F''(\rho)) \Delta_L(\rho) I_{12}^{(L)}(\rho) \nonumber \\
 - (N-1) \frac{1}{2 \rho} \bigl( \Delta_{21}(\rho)+\Delta_{11}(\rho) \bigr) J_{11}^{(T)}(\rho) \nonumber \\
- \frac{1}{2} \Delta_L'(\rho) J_{11}^{(L)}(\rho).
\label{RG_F2}
\end{eqnarray}
It can be easily checked that, in the equilibrium case ($v=0$), the equation can be reduced to that of the RF$O(N)$M, which is given in Ref.~\onlinecite{Tarjus-08}.

\section{RG equations for $X_k v_k$ and $Z_k$}
\label{sec:Appendix_RG_Z}

\subsection{Graphical representation for $\partial_l \Gamma_1^{(2)}$}

\begin{figure}
 \centering
 \includegraphics[width=0.45\textwidth]{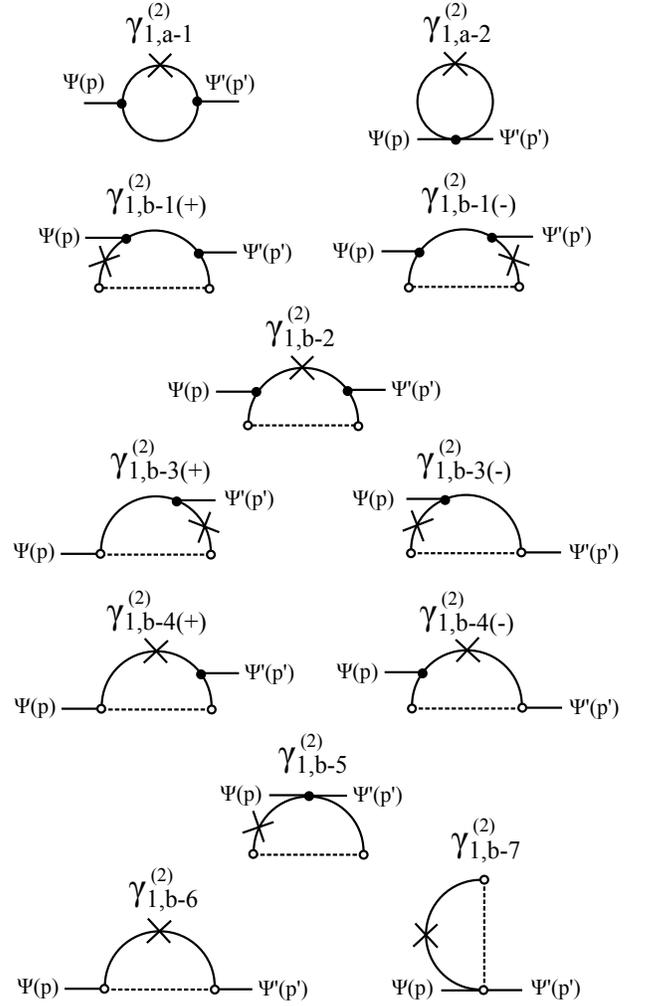}
 \caption{Graphical representation for the flow equation of $\Gamma_{1}^{(2)}$.}
 \label{fig:gamma_1-2}
\end{figure}

In this Appendix, we derive the RG equations for $X_k v_k$ and $Z_k$, which are given by Eqs.~(\ref{Def_v}) and (\ref{Def_Z}), respectively.
From Eq.~(\ref{RG_Gamma_1_graph}), we have the exact flow equation for $\Gamma_{1;\Psi(p),\Psi'(p')}^{(2)} = \delta^2 \Gamma_1/\delta \Psi(p) \delta \Psi'(p')$, where $\Psi$ represents $\psi^{\mu}$ or $\hat{\psi}^{\mu}$ and $p=(\mathbf{p},\omega_p)$, as follows:
\begin{eqnarray}
\partial_l \Gamma_{1;\Psi(p),\Psi'(p')}^{(2)} = \frac{1}{2} \bigl[ 2 \gamma_{1,a-1}^{(2)} - \gamma_{1,a-2}^{(2)} + 2 \gamma_{1,b-1(+)}^{(2)} \nonumber \\
+ 2 \gamma_{1,b-1(-)}^{(2)} +  2 \gamma_{1,b-2}^{(2)} -  2 \gamma_{1,b-3(+)}^{(2)} \nonumber \\
- 2 \gamma_{1,b-3(-)}^{(2)} -2 \gamma_{1,b-4(+)}^{(2)} -2 \gamma_{1,b-4(-)}^{(2)} \nonumber \\
- 2 \gamma_{1,b-5}^{(2)} + 2 \gamma_{1,b-6}^{(2)} + 2 \gamma_{1,b-7}^{(2)} \bigr].
\label{RG_Gamma_1-2}
\end{eqnarray}
The graphical representation of each term is shown in Fig.~\ref{fig:gamma_1-2}.
All functional derivatives are evaluated for a uniform field $ \bvec{\psi}_{rt} \equiv {}^t(\sqrt{2\rho_m},0,...,0)$, $\bvec{\hat{\psi}}_{rt} \equiv {}^t(0,...,0) $, where $\rho_m$ is the renormalized spontaneous magnetization defined by Eq.~(\ref{rho_m_def}).
We set $\Psi=\hat{\psi}^2$, $\Psi'=\psi^2$, and $p'=-p$ for the calculation of $\partial_l (X_k v_k)$ and $\partial_l Z_k$.
Note that $\gamma_{1,a-2}^{(2)}$, $\gamma_{1,b-5}^{(2)}$, $\gamma_{1,b-6}^{(2)}$, and $\gamma_{1,b-7}^{(2)}$ do not depend on the external momentum $\mathbf{p}$, thus they do not contribute to the RG equations for $X_k v_k$ and $Z_k$.

\subsection{Leading order contributions to $\partial_l X v$ and $\partial_l Z$}

As mentioned in Sec.~\ref{sec:RG_equation}, near the lower critical dimension $D=D_{\mathrm{lc}}+\epsilon$, $\gamma_{1}^{(2)}$ is expanded in terms of $\rho^{-1} \sim \epsilon$.
When $\gamma_1^{(2)}$ is expanded, we can see that $\gamma_{1,b-3}^{(2)}$ and $\gamma_{1,b-4}^{(2)}$ are sub-leading compared to $\gamma_{1,b-1}^{(2)}$ and $\gamma_{1,b-2}^{(2)}$.
Indeed, the former term's vertices yield a contribution of $2 \rho F'(\rho) \Delta'(\rho)$, while the latter term's vertices yield a contribution of $2 \rho F'(\rho)^2 \Delta(\rho)$, which is larger than that of the former if we assume that $\tilde{F}'(\tilde{\rho}_m)$ has a value of $\mathcal{O}(1)$ at the fixed point. 
Thus, we consider only $\gamma_{1,a-1}^{(2)}$, $\gamma_{1,b-1}^{(2)}$, and $\gamma_{1,b-2}^{(2)}$.

The term $\gamma_{1,a-1}^{(2)}$ is given by
\begin{eqnarray}
\gamma_{1,a-1}^{(2)} = 4 \rho_m F'(\rho_m)^2 X T \int_{\mathbf{q},\omega_q} \partial_l R_k(\mathbf{q}) \bigl[  \nonumber \\
 2 M_1(\mathbf{q}) D_1(\mathbf{q},\omega_q)^{-2} P_{12}^{(T)}(\mathbf{p+q},\omega_p+\omega_q) \nonumber \\
 + P_{21}^{(L)}(\mathbf{q},\omega_q)^2 D_0(\mathbf{p+q},\omega_p+\omega_q)^{-1} + (T \leftrightarrow L) \bigr],
 \label{gamma_1-2_a-1}
\end{eqnarray}
where $M(\mathbf{q})$, $D(\mathbf{q},\omega_q)$, and $P(\mathbf{q},\omega_q)$ are defined in Appendix \ref{sec:Appendix_Green_function}.
The symbol $(T \leftrightarrow L)$ means that $M_0$, $D_0$, and $P^{(T)}$ are exchanged by $M_1$, $D_1$, and $P^{(L)}$, respectively.
The term $\gamma_{1,a-1}^{(2)}$ is the same as the corresponding contribution in the pure $O(N)$ model because the velocity $v$ can be eliminated by changing the integral variable $\omega_q \to \omega_q + q_x v$.
Since we are interested in the zero temperature case, this term can be omitted.
$\gamma_{1,b-1}^{(2)}$, and $\gamma_{1,b-2}^{(2)}$ are given by
\begin{eqnarray}
\gamma_{1,b-1(\pm)}^{(2)} = 2 \rho_m F'(\rho_m)^2 \Delta_{00}(\rho_m) \int_{\mathbf{q}} \partial_l R_k(\mathbf{q})  \nonumber \\
\times P_{21}^{(T)}(\mathbf{q},0)^2 P_{12}^{(T)}(\mathbf{q},0) P_{12}^{(L)}(\mathbf{p \pm q},\omega_p),
\label{gamma_1-2_b-1}
\end{eqnarray}
\begin{eqnarray}
\gamma_{1,b-2}^{(2)} = 2 \rho_m F'(\rho_m)^2 \Delta_{00}(\rho_m) \int_{\mathbf{q}} \partial_l R_k(\mathbf{q})  \nonumber \\
\times P_{12}^{(T)}(\mathbf{-p+q},0) P_{21}^{(T)}(\mathbf{-p+q},0) P_{12}^{(L)}(\mathbf{q},\omega_p)^2,
\label{gamma_1-2_b-2}
\end{eqnarray}
where we have ignored the sub-leading terms of $\rho^{-1}$.
The propagators $P(\mathbf{q},\omega_q)$ are evaluated at $\rho=\rho_m$.
The momentum derivatives of Eqs.~(\ref{gamma_1-2_b-1}) and (\ref{gamma_1-2_b-2}) lead to the RG equations for $X_k v_k$ and $Z_k$.

\section{RG equation for $\Delta_k$}
\label{sec:Appendix_RG_Delta}

\subsection{Graphical representation for $\partial_l \Gamma_2^{(11)}$}

\begin{figure}
 \centering
 \includegraphics[width=0.3\textwidth]{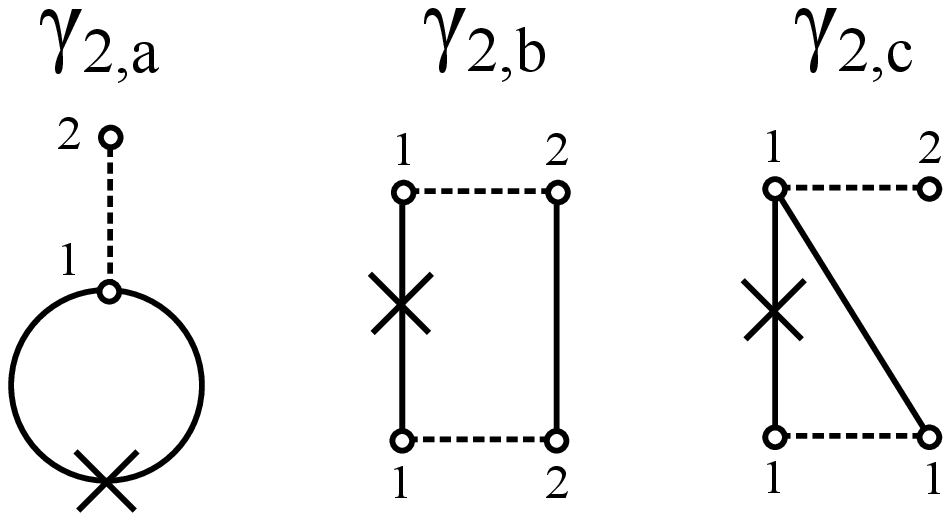}
 \caption{Graphical representation for the flow equation of $\Gamma_{2}$.}
 \label{fig:gamma_2}
\end{figure}

\begin{figure}
 \centering
 \includegraphics[width=0.45\textwidth]{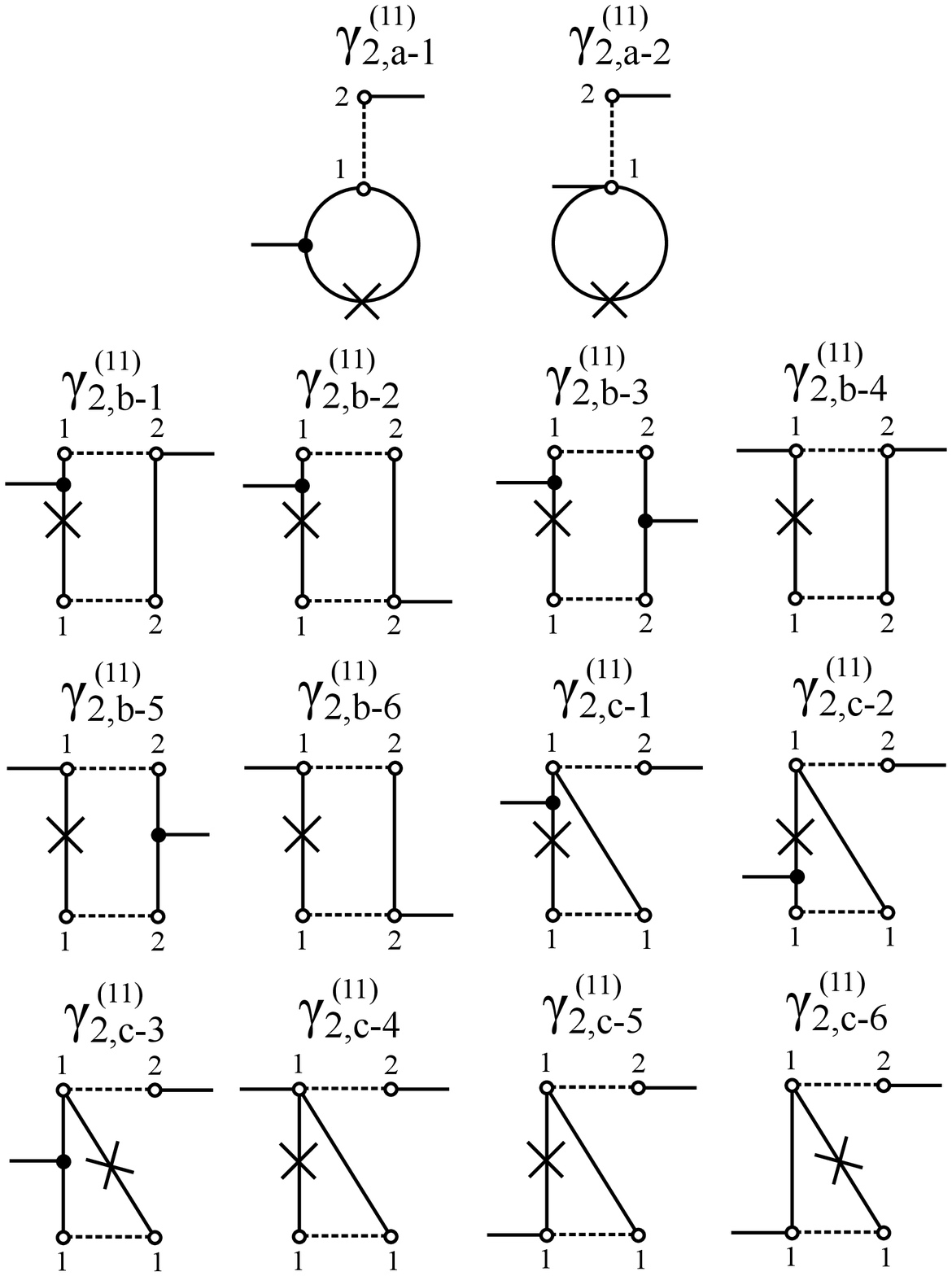}
 \caption{Graphical representation for the flow equation of $\Gamma_{2}^{(11)}$.}
 \label{fig:gamma_2-2}
\end{figure}

In this Appendix, we derive the RG equation for $\Delta_k(\bvec{\psi}_1,\bvec{\psi}_2)$, which is given by Eq.~(\ref{Def_Delta}).
To do this, the exact flow equation for $\Gamma_2^{(11)}$ is required.
Eq.~(\ref{RG_Gamma_2}) is rewritten as
\begin{equation}
\partial_l \Gamma_{2}[ \Psi_1, \Psi_2 ] = - \frac{1}{2} \bigl[ \gamma_{2,a} + \gamma_{2,b} + 2 \gamma_{2,c} + \mathrm{perm} \bigr],
\end{equation}
where $\gamma_{2,a}$, $\gamma_{2,b}$, and $\gamma_{2,c}$ are given in Fig.~\ref{fig:gamma_2} and ``perm'' denotes the permutation between the indices 1 and 2.
The flow equation for $\Gamma_{2;\hat{\psi}_1^{\mu} \hat{\psi}_2^{\nu}}^{(11)} = \delta^2 \Gamma_2/\delta \hat{\psi}_1^{\mu} \delta \hat{\psi}_2^{\nu}$ is then written as
\begin{eqnarray}
\partial_l \Gamma_{2;\hat{\psi}_1^{\mu} \hat{\psi}_2^{\nu}}^{(11)}[ \Psi_1, \Psi_2 ] = - \frac{1}{2} \bigl[ -2 \gamma_{2,a-1}^{(11)} + \gamma_{2,a-2}^{(11)} \nonumber \\
- 2 \gamma_{2,b-1}^{(11)} -2 \gamma_{2,b-2}^{(11)} + 2 \gamma_{2,b-3}^{(11)} + 2 \gamma_{2,b-4}^{(11)} \nonumber \\
-2 \gamma_{2,b-5}^{(11)} + 2 \gamma_{2,b-6}^{(11)} - 2 \gamma_{2,c-1}^{(11)} -2 \gamma_{2,c-2}^{(11)} \nonumber \\
- 2 \gamma_{2,c-3}^{(11)} + 2 \gamma_{2,c-4}^{(11)} + 2 \gamma_{2,c-5}^{(11)} + 2 \gamma_{2,c-6}^{(11)} +\mathrm{perm} \bigr],
\label{RG_Gamma_2-2}
\end{eqnarray}
where $\gamma_{2,a-1}^{(11)},...,\gamma_{2,c-6}^{(11)}$ are shown in Fig.~\ref{fig:gamma_2-2} and ``perm'' denotes the permutation between the indices 1 and 2, $\mu$ and $\nu$.
For example, $\gamma_{2,b-1}^{(11)}$ is written as
\begin{eqnarray}
\gamma_{2,b-1}^{(11)} = \mathrm{Tr} \bigl[ \partial_l \bvec{\mathrm{R}}_k(\mathbf{q}) \mathrm{P}_k[\Psi_1] \Gamma_{1;\hat{\psi}_1^{\mu}}^{(3)}[\Psi_1] \mathrm{P}_k[\Psi_1] \nonumber \\
\times \Gamma_{2;\hat{\psi}_2^{\nu}}^{(12)}[\Psi_1,\Psi_2] \mathrm{P}_k[\Psi_2] \Gamma_{2}^{(11)}[\Psi_2,\Psi_1] \mathrm{P}_k[\Psi_1] \bigr].
\label{gamma_2_b-1}
\end{eqnarray}
All functional derivatives are evaluated for a uniform field $ \bvec{\psi}_{1,rt} \equiv \bvec{\psi}_1$, $\bvec{\hat{\psi}}_{1,rt} \equiv \bvec{0} $, $ \bvec{\psi}_{2,rt} \equiv \bvec{\psi}_2$, and $\bvec{\hat{\psi}}_{2,rt} \equiv \bvec{0} $.

The functional derivatives of $\Gamma_1[\Psi]$ and $\Gamma_2[\Psi_1,\Psi_2]$ are calculated as
\begin{eqnarray}
\Gamma_{1; \psi^{\alpha} \psi^{\beta} \hat{\psi}^{\mu}}^{(3)}(q_1,q_2,q_3) = \bigl\{ -F''(\rho) \psi^{\alpha} \psi^{\beta} \psi^{\mu} \nonumber \\
-F'(\rho) (\delta^{\alpha \beta} \psi^{\mu} + \delta^{\alpha \mu} \psi^{\beta} + \delta^{\beta \mu} \psi^{\alpha}) 
 \bigr\} \nonumber \\
 \times (2\pi)^{D+1}\delta(\mathbf{q}_1+\mathbf{q}_2+\mathbf{q}_2)\delta(\omega_1+\omega_2+\omega_2),  \nonumber
\end{eqnarray}
\begin{eqnarray}
\Gamma_{2; \hat{\psi}_1^{\mu} \hat{\psi}_2^{\nu}}^{(11)}(q_1,q_2) = \Delta^{\mu \nu}(\bvec{\psi}_1,\bvec{\psi}_2) \nonumber \\
\times (2\pi)^{D+2} \delta(\mathbf{q}_1+\mathbf{q}_2)  \delta(\omega_1) \delta(\omega_2), \nonumber
\end{eqnarray}
\begin{eqnarray}
\Gamma_{2; \psi_1^{\alpha} \hat{\psi}_1^{\beta} \hat{\psi}_2^{\mu}}^{(21)}(q_1,q_2,q_3) = \partial_{\psi_1^{\alpha}} \Delta^{\beta \mu}(\bvec{\psi}_1,\bvec{\psi}_2) \nonumber \\
\times (2\pi)^{D+2}\delta(\mathbf{q}_1+\mathbf{q}_2+\mathbf{q}_3) \delta(\omega_1+\omega_2) \delta(\omega_3), \nonumber 
\end{eqnarray}
\begin{eqnarray}
\Gamma_{2; \psi_1^{\alpha} \psi_1^{\beta} \hat{\psi}_1^{\mu} \hat{\psi}_2^{\nu}}^{(31)}(q_1,q_2,q_3,q_4) = \partial_{\psi_1^{\alpha}} \partial_{\psi_1^{\beta}} \Delta^{\mu \nu}(\bvec{\psi}_1,\bvec{\psi}_2) \nonumber \\ 
\times (2\pi)^{D+2}\delta(\mathbf{q}_1+\mathbf{q}_2+\mathbf{q}_3+\mathbf{q}_4) \nonumber \\
\times \delta(\omega_1+\omega_2+\omega_3) \delta(\omega_4), \nonumber 
\end{eqnarray}
\begin{eqnarray}
\Gamma_{2; \psi_1^{\alpha} \hat{\psi}_1^{\mu} \psi_2^{\beta} \hat{\psi}_2^{\nu}}^{(22)}(q_1,q_2,q_3,q_4) = \partial_{\psi_1^{\alpha}} \partial_{\psi_2^{\beta}} \Delta^{\mu \nu}(\bvec{\psi}_1,\bvec{\psi}_2) \nonumber \\ 
\times (2\pi)^{D+2}\delta(\mathbf{q}_1+\mathbf{q}_2+\mathbf{q}_3+\mathbf{q}_4) \nonumber \\
\times \delta(\omega_1+\omega_2) \delta(\omega_3+\omega_4), \nonumber 
\end{eqnarray}
where $\Delta^{\mu \nu}(\bvec{\psi}_1,\bvec{\psi}_2)$ is expressed as Eq.~(\ref{Delta_expansion}).

\subsection{Leading order contribution to $\partial_l \Delta$}

As mentioned in Sec.~\ref{sec:RG_equation}, near the lower critical dimension $D=D_{\mathrm{lc}}+\epsilon$, $\gamma_{2}^{(11)}$ is expanded in terms of $\rho^{-1} \sim \epsilon$.
In addition, $\gamma_{2}^{(11)}$ is rewritten as
\begin{eqnarray}
\gamma_{2}^{(11)} = \gamma_{2,00}^{(11)} \delta^{\mu \nu} + (4 \rho_1 \rho_2)^{-1/2} \bigl[ \gamma_{2,12}^{(11)} \psi_1^{\mu} \psi_2^{\nu} \nonumber \\
+ \gamma_{2,21}^{(11)} \psi_2^{\mu} \psi_1^{\nu} + \gamma_{2,11}^{(11)} \psi_1^{\mu} \psi_1^{\nu} + \gamma_{2,22}^{(11)} \psi_2^{\mu} \psi_2^{\nu} \bigr].
\label{gamma_expansion}
\end{eqnarray}
Each coefficient of Eq.~(\ref{gamma_expansion}) can be calculated in the leading order of $\rho^{-1}$.
For example, $\gamma_{2,b-1,00}^{(11)}$ and $\gamma_{2,b-1,21}^{(11)}$ are given by
\begin{eqnarray}
\gamma_{2,b-1,00}^{(11)} = - (4 \rho_1 \rho_2)^{-1/2} \bigl( \Delta_{12} + \sqrt{\rho_2/\rho_1} z \Delta_{22} \bigr) \nonumber \\
\times \Delta_{00} I_{21}^{(T)}(\rho_1,\rho_2), \nonumber 
\end{eqnarray}
\begin{eqnarray}
\gamma_{2,b-1,21}^{(11)} = - (4 \rho_1 \rho_2)^{-1/2} \bigl[ 
\bigl\{ - z \Delta_{00} + (1-z^2) \Delta_{21} \bigr\} \nonumber \\
\times ( \partial_z \Delta_{00} + \Delta_{21} + z \partial_z \Delta_{21} + \sqrt{\rho_1/\rho_2} \partial_z \Delta_{11} ) \nonumber \\
+ \Delta_{21} ( \Delta_{12} + \sqrt{\rho_2/\rho_1} z \Delta_{22} ) \bigr] I_{21}^{(T)}(\rho_1,\rho_2). \nonumber
\end{eqnarray}
The function $I$ is defined by Eqs.~(\ref{I_def}) in Appendix \ref{sec:Appendix_Green_function}.
We have used $\psi^{\alpha} P_{12}^{\alpha \beta}=P_{12}^{(L)} \psi^{\beta} \simeq -(2\rho F'(\rho))^{-1} \psi^{\beta}$ and $\psi^{\alpha} P_{11}^{\alpha \beta}=P_{11}^{(L)} \psi^{\beta} \simeq 2 X T (2\rho F'(\rho))^{-2} \psi^{\beta}$, where the propagators $P_{ij}$ are also defined in Appendix \ref{sec:Appendix_Green_function}.
The other terms are straightforward to calculate.
$\gamma_{2,00}^{(11)}$ and $\gamma_{2,21}^{(11)}$ yield the RG equations for $\Delta_{00}$ and $\Delta_{21}$, respectively.
In the leading order of $\rho^{-1}$, $\Delta_{12}$, $\Delta_{11}$, and $\Delta_{22}$ do not appear in $\gamma_{2,00}^{(11)}$ and $\gamma_{2,21}^{(11)}$.
Thus, the RG equations for $\Delta_{00}$ and $\Delta_{21}$ compose a closed set of equations.

\section{Propagators}
\label{sec:Appendix_Green_function}

In this Appendix, we show the expression for the one-replica propagator Eq.~(\ref{Green_function}).
The functional derivative $\Gamma_{1,k}^{(2)}[\Psi]$ is evaluated for a uniform field $\bvec{\psi}_{rt} \equiv {}^t(\psi^1,...,\psi^N)$ and $\bvec{\hat{\psi}}_{rt} \equiv \bvec{0}$.
$\Gamma_{1,k}[\Psi]$ is given by Eq.~(\ref{Gamma_1}).
For simplicity, we omit the subscript $k$ in the following.
We introduce $P(q;\bvec{\psi})$ as
\begin{eqnarray}
\mathrm{P}[\Psi]_{q_1,q_2} = P(q_1;\bvec{\psi}) (2\pi)^{D+1} \delta(q_1+q_2).
\end{eqnarray}
where $q=(\mathbf{q},\omega)$ and $\delta(q)=\delta(\mathbf{q})\delta(\omega)$.
$P(q;\bvec{\psi})$ is a $2N \times 2N$ matrix, thus we write its element as $P_{ij}^{\mu \nu}$, where $i,j=1,2$ represent the two conjugate fields $\psi$ and $\hat{\psi}$, and $\mu, \nu=1,...,N$ are the field component indices.

$P_{ij}^{\mu \nu}(q;\bvec{\psi})=P_{ij}^{\mu \nu}(\mathbf{q},\omega;\bvec{\psi})$ can be written as 
\begin{eqnarray}
P_{ij}^{\mu \nu}(\mathbf{q},\omega;\bvec{\psi}) = P^{(T)}_{ij}(\mathbf{q},\omega;\rho) \biggl( \delta^{\mu \nu} - \frac{ \psi^{\mu} \psi^{\nu}}{2\rho} \biggr) \nonumber \\
+  P^{(L)}_{ij}(\mathbf{q},\omega;\rho) \frac{ \psi^{\mu} \psi^{\nu}}{2\rho},
\end{eqnarray}
where the transverse and longitudinal parts are given by
\begin{eqnarray}
P^{(T)}_{11}(\mathbf{q},\omega;\rho) &=& \frac{2 X T} {D_0(\mathbf{q},\omega;\rho)}, \nonumber \\ 
P^{(T)}_{12}(\mathbf{q},\omega;\rho) &=& \frac{M_0(\mathbf{q};\rho)-i(\omega-q_x v)X}{D_0(\mathbf{q},\omega;\rho)}, \nonumber \\
P^{(T)}_{21}(\mathbf{q},\omega;\rho) &=& \frac{M_0(\mathbf{q};\rho)+i(\omega-q_x v)X}{D_0(\mathbf{q},\omega;\rho)}, \nonumber \\
P^{(T)}_{22}(\mathbf{q},\omega;\rho) &=& 0,
\end{eqnarray}
\begin{eqnarray}
P^{(L)}_{11}(\mathbf{q},\omega;\rho) &=& \frac{2 X T}{D_1(\mathbf{q},\omega;\rho)}, \nonumber \\
P^{(L)}_{12}(\mathbf{q},\omega;\rho) &=& \frac{M_1(\mathbf{q};\rho)-i(\omega-q_x v)X}{D_1(\mathbf{q},\omega;\rho)}, \nonumber \\
P^{(L)}_{21}(\mathbf{q},\omega;\rho) &=& \frac{M_1(\mathbf{q};\rho)+i(\omega-q_x v)X}{D_1(\mathbf{q},\omega;\rho)}, \nonumber \\
P^{(L)}_{22}(\mathbf{q},\omega;\rho) &=& 0.
\end{eqnarray}
$M(\mathbf{q};\rho)$ and $D(\mathbf{q},\omega;\rho)$ are defined as
\begin{eqnarray}
M_0(\mathbf{q};\rho) &=& Z_{\parallel} q_x^2 + Z_{\perp} q_{\perp}^2 + R_k(\mathbf{q}) - F(\rho), \nonumber \\
M_1(\mathbf{q};\rho) &=& Z_{\parallel} q_x^2 + Z_{\perp} q_{\perp}^2 + R_k(\mathbf{q}) - F(\rho) - 2\rho F'(\rho), \nonumber \\
D_0(\mathbf{q},\omega;\rho) &=& M_0(\mathbf{q};\rho)^2 + ( \omega - q_x v )^2 X^2, \nonumber \\
D_1(\mathbf{q},\omega;\rho) &=& M_1(\mathbf{q};\rho)^2 + ( \omega - q_x v )^2 X^2. 
\label{M_D_def}
\end{eqnarray}
In Sec.~\ref{sec:RG_equation}, we also use the simplified notation $D(\mathbf{q};\rho)=D(\mathbf{q},\omega=0;\rho)$.

To express the RG equations in a compact form, we introduce the following integrals:
\begin{eqnarray}
L_{n}^{(T)}(\rho) = - \int_{\mathbf{q}} \partial_l R_k(\mathbf{q}) M_0(\mathbf{q};\rho)^{-n},
\label{L_def}
\end{eqnarray}
\begin{eqnarray}
I_{n n'}^{(T)}(\rho_1,\rho_2)  \nonumber \\
= - \int_{\mathbf{q}} \partial_l R_k(\mathbf{q}) P^{(T)}_{21}(\mathbf{q};\rho_1)^{n} P^{(T)}_{12}(\mathbf{q};\rho_2)^{n'},
\label{I_def}
\end{eqnarray}
\begin{eqnarray}
J_{n n'}^{(T)}(\rho_1,\rho_2) \nonumber \\
= - \int_{\mathbf{q}} \partial_l R_k(\mathbf{q}) P^{(T)}_{21}(\mathbf{q};\rho_1)^{n} P^{(T)}_{21}(\mathbf{q};\rho_2)^{n'},
\label{J_def}
\end{eqnarray}
where $\partial_l=-k\partial_k$, $\int_{\mathbf{q}} = \int d^D \mathbf{q}/(2\pi)^D$, and all frequencies $\omega$ in $P^{(T)}(\mathbf{q};\rho)$ are set to zero.
$L_n^{(L)}$, $I_{nn'}^{(L)}$, and $J_{nn'}^{(L)}$ are also defined by replacing $M_0$ and $D_0$ in Eqs.~(\ref{L_def})--(\ref{J_def}) with $M_1$ and $D_1$, respectively.
These integrals are used in Eqs.~(\ref{RG_rho_m}), (\ref{RG_Delta00}), and (\ref{RG_Delta21}).

\section{Numerical method to obtain non-analytic fixed functions}
\label{sec:Appendix_numerical}

In this Appendix, the numerical method used to obtain the fixed functions $\delta_{00}^*(z)$ and $\delta_{21}^*(z)$ is presented.
Since the solution exhibits a non-analytic behavior near $z=1$, standard numerical techniques are not applicable.
We define $\tilde{\delta}_{00}(z)=(N-2)\delta_{00}(z)$ and $\tilde{\delta}_{21}(z)=(N-2)\delta_{21}(z)$, and the trial functions $\tilde{\delta}_{00}^{(\mathrm{t})}(z)$ and $\tilde{\delta}_{21}^{(\mathrm{t})}(z)$ as follows,
\begin{eqnarray}
\tilde{\delta}_{00}^{(\mathrm{t})}(z) = a_0 + \sum_{n=1}^{n_{\mathrm{max}}} a_n (1-z)^{n/2},
\end{eqnarray}
\begin{eqnarray}
\tilde{\delta}_{21}^{(\mathrm{t})}(z) = b_0 + \sum_{n=1}^{n_{\mathrm{max}}} b_n (1-z)^{n/2}.
\label{delta_trial}
\end{eqnarray}
We rewrite Eqs.~(\ref{RG_delta00}) and (\ref{RG_delta21}) as
\begin{eqnarray}
\partial_l \delta_{00}(z) &=& \beta_{00} \bigl[ \delta_{00}, \delta_{21}; \epsilon \bigr](z), \nonumber \\
\partial_l \delta_{21}(z) &=& \beta_{21} \bigl[ \delta_{00}, \delta_{21}; \epsilon \bigr](z).
\end{eqnarray}
If $\epsilon$ is set to unity, the fixed functions $\tilde{\delta}_{00}^*(z)$ and $\tilde{\delta}_{21}^*(z)$ satisfy $\beta_{00} \bigl[ \tilde{\delta}_{00}^*, \tilde{\delta}_{21}^*; N-2 \bigr](z)=\beta_{21} \bigl[ \tilde{\delta}_{00}^*, \tilde{\delta}_{21}^*; N-2 \bigr](z)=0$.
The integral $S(\{ a_n \},\{ b_n \})$ is introduced,
\begin{eqnarray}
S(\{ a_n \},\{ b_n \}) = \int_{-1}^{1} \Bigl\{ \beta_{00}\bigl[ \tilde{\delta}_{00}^{(\mathrm{t})}, \tilde{\delta}_{21}^{(\mathrm{t})}; N-2 \bigr](z)^2 \nonumber \\
+ \beta_{21}\bigl[ \tilde{\delta}_{00}^{(\mathrm{t})}, \tilde{\delta}_{21}^{(\mathrm{t})}; N-2 \bigr](z)^2 \Bigr\} \mathrm{d}z,
\end{eqnarray}
which vanishes if the true fixed functions are attained.
The set of optimal parameters $\{ a_n \}$ and $\{ b_n \}$ can be obtained by minimizing $S(\{ a_n \},\{ b_n \})$.
From Eq.~(\ref{RG_a0}), $a_0$ and $a_1$ satisfy $-a_0 + a_0^2 - (1/2) a_1^2 = 0$.
Since $a_0=1$ when $a_1=0$, we obtain the following constraint,
\begin{equation}
a_0 = \frac{1}{2} \Bigl( \: 1+\sqrt{1+2a_1^2} \: \Bigr),
\end{equation}
which enables us to avoid the trivial solution $\{ a_n \} = \{ b_n \} = 0$.
Note that the integral $S(\{ a_n \},\{ b_n \})$ has several local minima.
One is chosen such that it recovers the fixed function Eq.~(\ref{FP-KT}) at $N=2$.
The truncation number is fixed at $n_{\mathrm{max}}=4$.
The inclusion of the higher order terms only changes $\eta_{\perp}=\delta_{00}^*(1)$ by less than one percent.

By employing a similar method, the anomalous dimension $\eta$ for the RF$O(N)$M can be also calculated from Eq.~(\ref{RG_R_eq}).
We have checked that it agrees with the known value given in Refs.~\onlinecite{Feldman-02} and \onlinecite{Feldman-00}.


\end{document}